\documentclass[twocolumn]{aastex61}

\usepackage{amsmath}

\usepackage{ulem,soul}

\definecolor{dark-gray}{gray}{0.33}
\definecolor{dark-red}{rgb}{0.75, 0.00, 0.00}
\definecolor{hlcolor}{rgb}{0.95, 0.95, 0.85}
\sethlcolor{hlcolor}

\renewcommand\emph[1]{\textit{#1}}

\newcommand\cm{\,\rm cm}

\newcommand\g{\,\rm g}

\newcommand\Sc{\mathrm{Sc}}
\newcommand\St{\mathrm{St}}

\newcommand{\simgt}%
           {\,\hbox{\lower0.35ex\hbox{$\sim$}\llap{\raise0.35ex\hbox{$>$}}}\,}
\newcommand{\simlt}%
           {\,\hbox{\lower0.35ex\hbox{$\sim$}\llap{\raise0.35ex\hbox{$<$}}}\,}

\begin{document}

\title[Permeability of planet-induced gaps]{Characterizing the variable dust permeability of planet-induced gaps}
\author{Philipp Weber{$^\star$}}
\author{Pablo Ben{\'i}tez-Llambay}
\author{Oliver Gressel}
\author{Leonardo Krapp}
\author{Martin E. Pessah}

\affiliation{Niels Bohr International Academy, The Niels Bohr Institute, Blegdamsvej 17, DK-2100, Copenhagen \O, Denmark}

\begin{abstract}
Aerodynamic theory predicts that dust grains in protoplanetary disks will drift radially inward on comparatively short timescales. In this context, it has long been known that the presence of a gap opened by a planet can alter the dust dynamics significantly.
In this paper, we carry out a systematic study employing long-term numerical simulations aimed at characterizing the critical particle-size for retention outside a gap as a function of particle size and for various key parameters defining the protoplanetary disk model.
To this end, we perform multifluid hydrodynamical simulations in two dimensions, including different dust species, which we treat as pressureless fluids. We initialize the dust outside of the planet's orbit and study under which conditions dust grains are able to cross the gap carved by the planet.
In agreement with previous work, we find that the permeability of the gap depends both on dust dynamical properties and the gas disk structure: while small dust follows the viscously accreting gas through the gap, dust grains approaching a critical size are progressively filtered out.
Moreover, we introduce and compute a depletion factor that enables us to quantify the way in which higher viscosity, smaller planet mass, or a more massive disk can shift this critical size to larger values.
Our results indicate that gap-opening planets may act to deplete the inner reaches of protoplanetary disks of large dust grains -- potentially limiting the accretion of solids onto forming terrestrial planets.
\end{abstract}

\email{$\star$ philipp.weber@nbi.ku.dk}


\section{Introduction}

Gaseous protoplanetary disks (PPDs) are the sites where planets form. They comprise a class of intriguing, dynamically evolving astrophysical objects with a broad range of chemical and microphysical processes at play. Observationally, many characteristics of nearby PPDs such as, for instance, their near-Keplerian rotation, their radial surface brightness profiles, or the accretion rate onto the central star can now be determined with a reasonable degree of confidence
\citep[see][for a comprehensive review]{2011ARA&A..49...67W}.

Apart from their gas content and entrained magnetic flux, PPDs are believed to inherit the canonical interstellar dust-to-gas mass ratio of about one percent from their parental molecular cloud \citep{2011ARA&A..49...67W}. Moreover, from the observed frequency of PPDs in young stellar clusters, a median disk lifetime of roughly three to five million years is derived \citep[e.g.][]{2009AIPC.1158....3M,Alexander2014}. For planetesimals and ultimately protoplanets to form out of the initially minuscule interstellar dust grains, many orders of magnitudes in size (and mass) have to be overcome -- which already suggests that a combination of different growth/accumulation processes is required.

Overall, the dust mass in a young PPD is expected to be much smaller than the mass of the gas \citep{William2014}, but understanding the dynamics of the dust is nevertheless of paramount importance. Firstly, at the low temperature of the system, it is mainly the dust-thermal continuum emission that is observed by modern telescopes in the infrared and (sub-)millimeter bands. Secondly, evolved dust provides the materials from which planetesimals and ultimately planets are formed \citep{2014prpl.conf..339T}.

At the same time, the existence of gas giants puts imperative constraints on the timespan in which planet formation has to take place. Naturally, planets with a significant hydrogen content require gas to be present in the PPD at their time of formation \citep{1996Icar..124...62P}. In a gaseous disk however, a giant planet has a perturbing effect, depleting its orbit of material and by that creating a gap in the disk \citep{Lin1979}. Yet, in a viscously evolving disk, the gap cannot be entirely cleared of gas, and there remains an accretion flow through it \citep[e.g.,][]{1994ApJ...421..651A,Lubow2006}. It is well established that small enough grains couple tightly to the motion of the gas  \citep{Whipple1972, Weidenschilling1977}, and thus if these small grains are located outside of the planet's orbit, they simply follow the accreting gas through the planetary gap. In contrast to this, larger grains decouple from the gas and remain in the outer system.

In general, the behavior of dust grains is governed by its relative velocity with respect to the underlying gas disk, which in turn is strongly linked to local variations in the azimuthal velocity and ultimately to the pressure structure of the disk. Here, the presence of a gap and subsequently a pressure maximum just outside of the gap has the effect that dust grains, depending on their size, can become trapped \citep{Whipple1972}. High resolution imaging in recent years revealed some structures in protoplanetary disks, such as concentric rings that were most prominently observed in HL Tau \citep{ALMA2015,2015A&A...584A.110P,2016ApJ...818...76J} and TW Hydra \citep{Andrews2016,vanBoekel2017}. Such concentric ring structures are often, although not exclusively, ascribed to the presence of embedded massive planets \citep[e.g.,][]{Gonzalez2015,Bae2017}.

In the regime of very low levels of turbulence in the disk\footnote{This scenario is motivated by simulations including detailed treatment of non-ideal MHD effects \citep[see, e.g.,][]{Gressel2015}.}, recent simulations by \citet{Dong2017} and \citet{Bae2017} have shown that a single intermediate-mass planet can create secondary and even tertiary spiral arms that can open gaps of their own, which may then act as dust traps. The level of turbulence assumed in our present study is typically above the limits where secondary arms are produced \cite[cf. sect. 6 of][]{Bae2017}. Moreover, while this effect is certainly useful for interpreting systems of multiple dust rings, such as the ones observed in HL~Tau, overall the edge of the primary gap will likely retain the dominant role in preventing grains from reaching the inner disk.

The size-dependence of the dust trapping has been the subject of several numerical studies. An early investigation by \citet{Paardekooper2006} looked into the evolution of dust inside a planetary gap and the accretion flow of dust and gas onto the planet. Based on a two-dimensional simulation, \citet{Rice2006} used the azimuthally averaged gas density profile to establish an analytical expression for the radial gap structure of the gas, and its influence onto the dust dynamics. In more recent models, the effect of an embedded planet on the dust has been studied at increased numerical resolution. \cite{Pinilla2012} regarded how the pressure trap can be a preferred site for dust coagulation -- their study, as well as those of \cite{Zhu2012} and \cite{Owen2014} were interested in connecting the scenario to observations of so-called transition disks, that is, systems that show an inner cavity in the observed flux. Similar results regarding particle trapping at the outer gap edge were found for the case of meter-sized objects \citep{Ayliffe2012}, however, only for gaps opened by a relatively massive perturber. Local accumulation of pebbles and boulders in this size range may potentially lead to the rapid formation of planetesimals outside the orbit of an existing planet.

While previous studies have especially focused on the final (dust-)particle density structure in the trapped regions, we primarily turn our attention to the gap as a ``particle filter'' and to investigate on which parameters its permeability depends. By this, we aim to answer for which dust species the presence of the planet would create two radially separated reservoirs in a quantitative manner. This is motivated by recent analysis of chondritic material \citep[e.g.,][]{Olsen2016,Budde2016,Kruijer2017}, which suggests that, at some early point in the lifetime of the Solar System, there were two disconnected reservoirs of solids. In this work, we want to outline, how such a separation could have been created by an early formed Jupiter and how it would have influenced the size distributions of solids inside and outside of the planet's orbit. From this, we obtain a quantity which is possibly observable in chondrites and could constrain the physical conditions in the early Solar System.

Our paper is organized as follows: In Section~\ref{sec:model} we briefly describe the governing equations for gas and dust, in Section~\ref{sec:Setup} we introduce the reader to the model and the numerical implementation, we present the results in Section~\ref{sec:results} and discuss potential consequences in Section~\ref{sec:discussion}, before concluding in Section~\ref{sec:conclusion}.

\section{Disk Model}\label{sec:model} 

We aim to investigate in detail the processes by which dust is able to overcome a planet-carved gap. To this end, we formulate the simple question of how much dust from the outer system is allowed to cross the planet's orbit towards the inner disk and, conversely, how much of the dust is ``filtered out'' at the outer gap edge. Specifically, we keep the planet on a fixed circular orbit, treat the dust as a pressureless fluid (strictly valid only for $\St\ll 1$), and use an effective $\alpha$ viscosity. Turbulent dust diffusion and feedback of the dust fluids onto the gas are considered in special cases.

Assuming a viscous disk with an effective kinematic viscosity, $\nu$, the fundamental equations that define the evolution of the gas are the Navier-Stokes equations. We moreover deal with dust as an inviscid and pressureless fluid, described by the Euler equations.

Considering a two fluid system comprised of gas and a single dust species, the continuity equations for the gas and dust read
\begin{eqnarray}
\frac{\partial\Sigma_\mathrm{g}}{\partial t} + \mathbf{\nabla}\cdot \left(\Sigma_\mathrm{g}\mathbf{u} \right) & = & 0 \,, \label{eq:contgas} \\
\frac{\partial\Sigma_\mathrm{d}}{\partial t} + \mathbf{\nabla}\cdot \left(\Sigma_\mathrm{d}\mathbf{v} +\mathbf{j}\right) & = & 0 \,, \label{eq:contdust}
\end{eqnarray}
where $\Sigma_\mathrm{g}$ and $\Sigma_\mathrm{d}$ are the gas- and dust surface density, respectively, $\mathbf{u}$ and $\mathbf{v}$ are their respective velocities and $\mathbf{j}$ denotes a possible additional mass flux due to diffusion of dust particles. We adopt the model of \citet{Morfill1984}, in which $\mathbf{j}$ is given by:
\begin{equation}
\mathbf{j} = -D_{\rm d}\Sigma\, \mathbf{\nabla} \left( \frac{\Sigma_\mathrm{d}}{\Sigma}\right)\,,
\label{eq:diffflux}
\end{equation}
where $\Sigma$ is the combined gas and dust density, which -- under the assumption that $\Sigma_d \ll \Sigma_g$ -- is replaced by $\Sigma_g$. If we assume the gas diffusivity to be equal to the disk viscosity, then the diffusion coefficient $D_{\rm d}$ of the dust can be expressed by the so-called Schmidt number
\begin{equation}
\Sc \equiv \frac{D_{\rm g}}{D_{\rm d}}= \frac{\nu}{D_{\rm d}}\,.
\end{equation}
\cite{Youdin2007} argue that the Schmidt number is a function of the Stokes number, $\mathrm{St}$, defined in Equation~(\ref{eq:Stokes}) below, a dimensionless parameter that quantifies the coupling between gas and dust. Specifically, they predict
\begin{equation}
\Sc \approx 1+\St^2\,.
\end{equation}
Since in this work we focus on Stokes numbers below unity we approximate this dependence as $\Sc=1$ whenever we include the diffusive flux in our simulations.

The momentum equations for the gas and dust are of the usual form, and can be written as
\begin{eqnarray}
\Sigma_\mathrm{g}\frac{\mathrm{D}\mathbf{u}}{\mathrm{D} t}
& = &  - \mathbf{\nabla}P - \mathbf{\nabla}\cdot\tau \; - \Sigma_\mathrm{g}\mathbf{\nabla \phi} - \Sigma_{\rm d} \mathbf{f}_{\rm d} \,,
\label{eq:NS-gas} \\[4pt]
\Sigma_\mathrm{d}\frac{\mathrm{D}\mathbf{v}}{\mathrm{D} t}\; & = & \; -\Sigma_\mathrm{d}\mathbf{\nabla \phi} + \Sigma_{\rm d} \mathbf{f}_{\rm d}\,,
\label{eq:NS-dust}
\end{eqnarray}
for the gas- and dust fluid, respectively. Here, we used the definition of the Lagrangian derivative, that is,
\begin{equation}
\frac{\mathrm{D}}{\mathrm{D} t} \equiv \frac{\partial}{\partial t} + \mathbf{u}\cdot \mathbf{\nabla}
\end{equation}
for $\mathbf{u}$ in Equation~(\ref{eq:NS-gas}), and a similar expression for $\mathbf{v}$ in Equation~(\ref{eq:NS-dust}). Moreover, $P$ is the gas pressure, $\phi$ is the gravitational potential and $\mathbf{f}_{\rm d}$ is a function that represents the interaction between gas and dust species via a drag-force. The latter term is defined in Equation~(\ref{eq:coupling}) and discussed in detail in Section~\ref{sec:dust_disk} below. Momentum conservation implies that this term must appear in Equations~(\ref{eq:NS-gas}) and (\ref{eq:NS-dust})  with opposite signs. In this work we assume a vertically isothermal disk for which the pressure and surface density are connected by the sound speed, $c_\mathrm{s}$:
\begin{equation}
P = c_\mathrm{s}^2\Sigma_\mathrm{g}\,.
\end{equation}
The viscous stress tensor is furthermore given by
\begin{equation}
\tau \equiv \Sigma_{\rm g} \nu \left[ \mathbf{\nabla} \mathbf{u} + (\mathbf{\nabla}\mathbf{u})^T - \frac{2}{3}(\mathbf{\nabla}\cdot \mathbf{u})\mathit{\mathbf{I}}\right]\,,
\end{equation}
where $\mathbf{I}$ is the identity matrix.

\subsection{Gas disk} 

Considering the above equations, we are looking for a steady-state solution of the gas disk. From Equation~(\ref{eq:contgas}) and using the classical result by \citet{Pringle1981} for the radial velocity of a viscous accretion flow, it follows that
\begin{equation}
u_r=-\frac{3\nu}{2r}\,.
\label{eq:vgr}
\end{equation}
In steady-state, the gas surface density is described by
\begin{equation}
\Sigma_\mathrm{g}=-\frac{\dot{M}}{3\pi\nu}\,,
\label{eq:sigma-g}
\end{equation}
with $\dot{M}$ being a constant which is equal to the mass flow through the disk and in general negative (signifying inward flow of mass).

Throughout our study, we adopt the $\alpha$-viscosity prescription \citep{Shakura1973}, in which the viscosity is parametrized by $\nu = \alpha c_\mathrm{s} H_\mathrm{g}$, where $\alpha$ is a constant dimensionless parameter, $H_\mathrm{g}$ the scale height of the gas disk, which can be written as $H_\mathrm{g}(r)=c_\mathrm{s}(r)/\Omega_\mathrm{K}(r)$, with $\Omega_\mathrm{K}(r)=\sqrt{GM_\ast/r^3}$ being the Keplerian frequency. $G$ is the gravitational constant and $M_\ast$ is the mass of the central star. We moreover assume a power-law dependence for the sound speed, which by Equation~(\ref{eq:sigma-g}) directly implies a power-law for the surface density of the gas,
\begin{equation}
c_\mathrm{s}^2(r) \propto r^{\bar{q}},\qquad
\Sigma_{\mathrm{g}}(r) \propto r^{\bar{p}}\,.
\end{equation}
From Equation~(\ref{eq:sigma-g}) one can further derive that the pressure also shows a power-law dependence in $r$ with exponent $\bar{p}+\bar{q}=-3/2$. For simplicity, we chose the commonly assumed case of a non-flaring disk, that is $\bar{q}=-1$, which fixes $\bar{p}=-1/2$. This is a somewhat shallower mass profile than obtained for a flaring, irradiated disk. However, due to the above constraint fixing the sum of the two exponents, the resulting pressure profile
\begin{equation}
 P(r) \propto r^{\bar{p}+\bar{q}} \equiv r^{-3/2} \,,
\end{equation}
is independent of $\bar{p}$ and $\bar{q}$. While we do not expect the local pressure structure near the planet to depend sensitively on the particular choice for the global power-law indices, relaxing this restriction offers a potentially fruitful future extension of the scope of our findings.

When only considering the gravitational interaction with the star, $\phi=\phi_\ast=-GM_\ast/r$, and with the definition of $\Omega_\mathrm{K}$, one finds $\nabla \phi=r\Omega_\mathrm{K}^2$. By neglecting the viscous term and the contribution of the coupling to the dust, the radial component of Equation~(\ref{eq:NS-gas}) yields
\begin{equation}
u_\varphi = v_\mathrm{K} \sqrt{1-\eta}\,,
\label{eq:gasazi}
\end{equation}
where $v_\mathrm{K}=\Omega_\mathrm{K} r$ is the Keplerian velocity and $\eta$ is a function that modifies the Keplerian rotation by taking into account the additional pressure support of the disk. With
\begin{equation}
\eta \equiv -\left(\frac{H_{\rm g}}{r}\right)^2\frac{\partial \log{P}}{\partial \log{r}}\,,
\end{equation}
$\eta$ is in general a positive quantity much smaller than unity meaning that the unperturbed disk rotates with a slightly sub-Keplerian velocity.

\subsection{Dust disk} \label{sec:dust_disk}

We consider spherical dust grains of radius $a$ and intrinsic material density $\rho_\mathrm{int}$ assumed equal to $3\,\mathrm{g}\,\mathrm{cm}^{-3}$, similar to the value for silicates given by \citet{Zhukovska2008}. To fully understand the dust dynamics, it is important to take a closer look at the coupling function $\mathbf{f}_{\rm d}$ in Equation~(\ref{eq:NS-dust}). It is typically adopted that
\begin{equation}
\mathbf{f}_{\rm d} = \frac{\Sigma_\mathrm{g}}{a \rho_\mathrm{int}}\frac{2} {\pi}\,\Omega_\mathrm{K}\;(\mathbf{u}-\mathbf{v})\,,
\label{eq:coupling}
\end{equation}
in the two-dimensional case \citep[][p.15]{Safronov1972}, and \cite{Whipple1972} argues for this to be valid in the case of the Epstein regime, that is, the situation where the mean-free-path of gas molecules is approximately larger than the particle radius. We make sure that this condition is satisfied when considering a typical model for the Solar System(that is, $\Sigma_\mathrm{g}=10^3\,\mathrm{g}\,\mathrm{cm}^{-2}$, $H=0.05\,\mathrm{AU}$). Accordingly, we find an upper limit for the particle radius of $a_\mathrm{max}\approx 15\,\mathrm{cm}$, that we will avoid to exceed.

Following usual convention, we introduce the aforementioned Stokes number which in our context is given by
\begin{equation}
\St \equiv \tau_{\rm stop}\Omega_{\rm K}= \frac{a \rho_\mathrm{int}}{\Sigma_\mathrm{g}}\frac{\pi}{2}\,,
\label{eq:Stokes}
\end{equation}
such that Equation~(\ref{eq:coupling}) becomes
\begin{equation}
\mathbf{f}_{\rm d} = \frac{\Omega_\mathrm{K}}{\St}\,(\mathbf{u}-\mathbf{v})\,.
\end{equation}
In the way it is described in Equation~(\ref{eq:NS-dust}), the dust is treated as a pressureless fluid, and \cite{Hersant2009} points out that this is only a valid approximation for $\St\simlt 1/2$.
Naturally, particles become trapped at the edge of the gap most efficiently when their Stokes number approaches or mildly exceeds unity. As such, our fluid-based approach should capture the relevant physics. We revisit this issue in our concluding discussion by assessing the obtained Stokes numbers.

When considering gas-dust interactions, the equilibrium solution departs from the Keplerian disk. Assuming that this deviation is small, \cite{Nakagawa1986} give the solution for the angular and radial velocities. Neglecting the feedback of the dust onto the gas, \cite{Takeuchi2002} give the simplified expressions:
\begin{eqnarray}
\displaystyle{ v_{r}} & = & \displaystyle{\frac{\St^{-1} u_{r} - \eta\, v_\mathrm{K}}{\St+\St^{-1}}}\,,
 \label{eq:vdr}\\[4pt]
\displaystyle{ v_{\varphi}} & = & \displaystyle{u_\varphi - \frac{1}{2}\St\, v_r}\,.
\label{eq:dustvphi}
\end{eqnarray}
The radial velocity of the dust has two different contributions: small dust, which is tightly coupled to the gas, that is, $\St^{-1}\gg 1$ in Equation~(\ref{eq:vdr}), essentially moves with the radial velocity, $u_{r}$, of the background flow  -- cf. Equation~(\ref{eq:vgr}). Dust that is not completely coupled to the gas moves with a slightly different azimuthal velocity, which can be seen from Equation~(\ref{eq:dustvphi}) and, in this way, experiences an acceleration or deceleration via the friction force.

\section{Numerical Setup} \label{sec:Setup} 

For our simulations, we use the public version of the code FARGO3D\footnote{\href{http://fargo.in2p3.fr}{http://fargo.in2p3.fr}} \citep{Beni2016}, which solves the magnetohydrodynamical equations using finite-difference upwind, dimensionally split methods, combined with the FARGO algorithm \citep{Masset2000} for the orbital
advection and a fifth order Runge-Kutta integrator for planetary orbits. One important feature is that FARGO3D can run on GPU clusters, making this code an excellent tool to study long-term simulations (thousands of orbits) in reasonable wall-clock time (typically, a few days). We have extended it for solving the multifluid equations with an implicit solver, which allows us to evolve the coupled gas- and dust dynamics (Benitez-Llambay et al., in prep) at reasonable computational cost. In this work, we present two-dimensional simulations on a polar grid of $N_\varphi \times N_r = 1024\times 512$ cells, which provides sufficient resolution to study the problem at hand. We have checked that changing the resolution does not affect the outcome of the simulations.

\subsection{Initial conditions}\label{sec:condinit} 

The natural units in which to treat this problem are the mass of the star, the radial position of the planet and its orbital frequency. The units that we use are listed in Table~\ref{tab:codeunits}. Throughout this work, the planet is kept on a fixed circular orbit at its initial radius.

\begin{table}
\centering
        \caption{\textrm{Units.}}
  \label{tab:codeunits}
\begin{tabular}{l c r}
  \hline
  Mass \qquad  \qquad & $[M]=$ &$M_\ast$  \\
  Length \qquad  \qquad & $[L]=$ &$r_{\rm P}$ \\
  Time \qquad  \qquad & $[T]=$ &$\Omega_\mathrm{K}^{-1}\left(r_{\rm P}\right)$ \\[0.1cm]
  \hline
\end{tabular}
\end{table}

We simulate the protoplanetary disk in a radial range of $r \in \left[0.25,3.0\right]$. The planet is modeled by the influence of its gravitational potential
\begin{equation}
\phi_\mathrm{P} = -\frac{G\,m_\mathrm{P}}{\left(|\mathbf{r}-\mathbf{r}_\mathrm{P}|^2+\epsilon^2\right)^{\frac{1}{2}}} + \frac{G\,m_\mathrm{P}}{r_\mathrm{P}^2}r\cos{\varphi}\,,
\label{eq:planetpot}
\end{equation}
where $m_\mathrm{P}$ and $\mathbf{r}_\mathrm{P}$ are the mass and position of the planet, respectively. The first term corresponds to the gravitational potential of the planet and the second one, known as the indirect term, corresponds to the acceleration experienced by the center of the frame of reference due to the presence of the planet. The parameter $\epsilon$ is the so-called smoothing length, which accounts for the vertical extent of the disk in the two-dimensional setting \citep{Masset2002, Muller2012}. We set this parameter to a fiducial value of $\epsilon = 0.6H_\mathrm{P}$, with $H_\mathrm{P}$ the vertical scale height of the gas disk at the planet's location. We have tested that setting this parameter to smaller values does not change the outcome.

There are several parameters that need to be specified to fully define our model. In the fiducial case, we choose parameters suitable for a disk that is still quite young. The complete set of parameters is shown in Table~\ref{tab:para}, where $\dot{M}_{\ast}$ describes the stellar accretion rate and $q = M_\mathrm{P}/M_\ast$ denotes the mass ratio between the planet and the central star.

Before introducing the dust into the simulation, we aim to ascertain that the gap in the gas disk has reached an equilibrium state. \cite{Kanagawa2016} showed that, while the \emph{depth} of the gap reaches an equilibrium already after about 1,000 planet orbits, it takes about the viscous timescale for the gap \emph{width} to approach a stationary value. We thus simulate the gas for 10,000 orbits (at the planet location) before the dust fluids are introduced into the model.

Because we are first and foremost interested in the efficiency of the dust filtration by the planetary gap for dust grains that are drifting inwards, we set up an initial dust reservoir \emph{outside of the gap}, that is to say
\begin{equation}
\Sigma_{\rm d}(r) = \left\{\begin{array}{ll}
                    10^{-20}  & \qquad\mbox{if } r<1.5 \\
                    \varepsilon\,\Sigma_g & \qquad\mbox{if } r\geq 1.5
                    \end{array}\right.\,,
\label{eq:dustinit}
\end{equation}
where $\varepsilon = \Sigma_\mathrm{d}/\Sigma_\mathrm{g}$ is the dust-to-gas ratio for one individual dust species. In the case that one neglects the feedback of the dust onto the gas -- this means neglecting the contribution of $\Sigma_{\rm d}\textbf{f}_{\rm d}$ in Equation~(\ref{eq:NS-gas}) -- the result is not dependent on $\varepsilon$. However, we here set it to the conventional value of 0.01. In our simulations, the dust velocities are initialized with the analytical solutions given by Equations~(\ref{eq:vdr}) and (\ref{eq:dustvphi}) above.

\begin{table}
\centering
        \caption{\textrm{Set of fiducial model parameters.}}
  \label{tab:para}
\begin{tabular}{l c r }
  \hline
  viscosity parameter \qquad  \qquad & $\alpha$   \qquad& $3\times10^{-3}$ \\
  surface density slope \qquad  \qquad &$\bar{p}$   \qquad& -0.5 \\
  temperature slope \qquad  \qquad & $\bar{q}$   \qquad& -1.0 \\
  aspect ratio \qquad  \qquad&$h$   \qquad& 0.05 \\
  mass ratio \qquad  \qquad &$q$  \qquad & $10^{-3}$ \\
  Diffusion \qquad  \qquad &  \quad \qquad & no \\
  Feedback \qquad  \qquad & \quad \qquad & no \\
  \hline
\end{tabular}
\end{table}
After introducing the dust, we simulate the gas and dust simultaneously for further 10,000 orbits. The reason we have to evolve the simulation for this many orbits becomes apparent when calculating the timescale on which gas accretes from outside of the gap location to the inner parts in an unperturbed disk:
\begin{eqnarray}
\tau_{\rm acc} &\equiv& \int_{r_\mathrm{i}}^{r_\mathrm{f}}u_r^{-1} {\rm d}r \nonumber\\
                                &=&-\frac{2 v_\mathrm{K}(r_0)}{3 \alpha c_{\mathrm{s},0}^2} \int_{r_\mathrm{i}}^{r_\mathrm{f}} \left(\frac{r}{r_0}\right)^{-\bar{q}-\frac{1}{2}}\, {\rm d}r \nonumber\\[4pt]
                &=& \frac{2 }{3 \alpha h^{2}(r_0)\gamma}\left[\left(\frac{r_\mathrm{i}}{r_0}\right)^{\gamma} - \left(\frac{r_\mathrm{f}}{r_0}\right)^{\gamma}\right]\Omega_\mathrm{K}^{-1}(r_0)\,,\qquad
\label{eq:timescale}
\end{eqnarray}
where we used $\gamma = (-\bar{q}+1/2)$ as a substitution.  Throughout this work we set $r_0 \equiv 1.0$. Demanding that the gas moves from an initial position at $r_\mathrm{i}=1.5$ to a final position at $r_\mathrm{f}=0.5$, one finds that the timescale for the fiducial model would be $\tau_{\rm acc}\approx 14,000$ orbits. In our simulations, however, we saw that with a massive planet being present, this process is sped up considerably.

\subsection{Boundary conditions}\label{sec:boundaries} 
In order to better characterize the permeability of the planetary gap, we prescribe inflow of mass at the outer boundary, that is, far enough away from the planet's position to be able to apply the equilibrium solutions we obtained in Section~\ref{sec:model} \citep[see also][]{Durmann2015}. Furthermore, it is necessary to provide separate boundary conditions for the gas and every individual dust component, as especially their radial velocities can be vastly different. Therefore, at the outer boundary, we specify the gas velocity and density according to Equations~(\ref{eq:vgr}) and (\ref{eq:sigma-g}) and the velocities and densities of the dust species are given by Equations~(\ref{eq:vdr}) and (\ref{eq:dustinit}), respectively. Also, the azimuthal velocities have to be fixed then to the values given by Equation~(\ref{eq:gasazi}) for the gas and by Equation~(\ref{eq:dustvphi}) for the dust. In order to minimize possible reflections of the planet's wake at the outer boundary, we moreover damp all fields consistently with the procedure described in \cite{deVal-Borro2006}. We do not fix the mass outflow at the inner radial boundary since it may differ from the equilibrium solution owing to the the presence of the planet. Instead, we only force the velocities to the unperturbed disk solution and implement a zero-gradient boundary for the gas- and dust densities.

\section{Results}\label{sec:results} 

\begin{figure}
\centering
\includegraphics[scale=0.5]{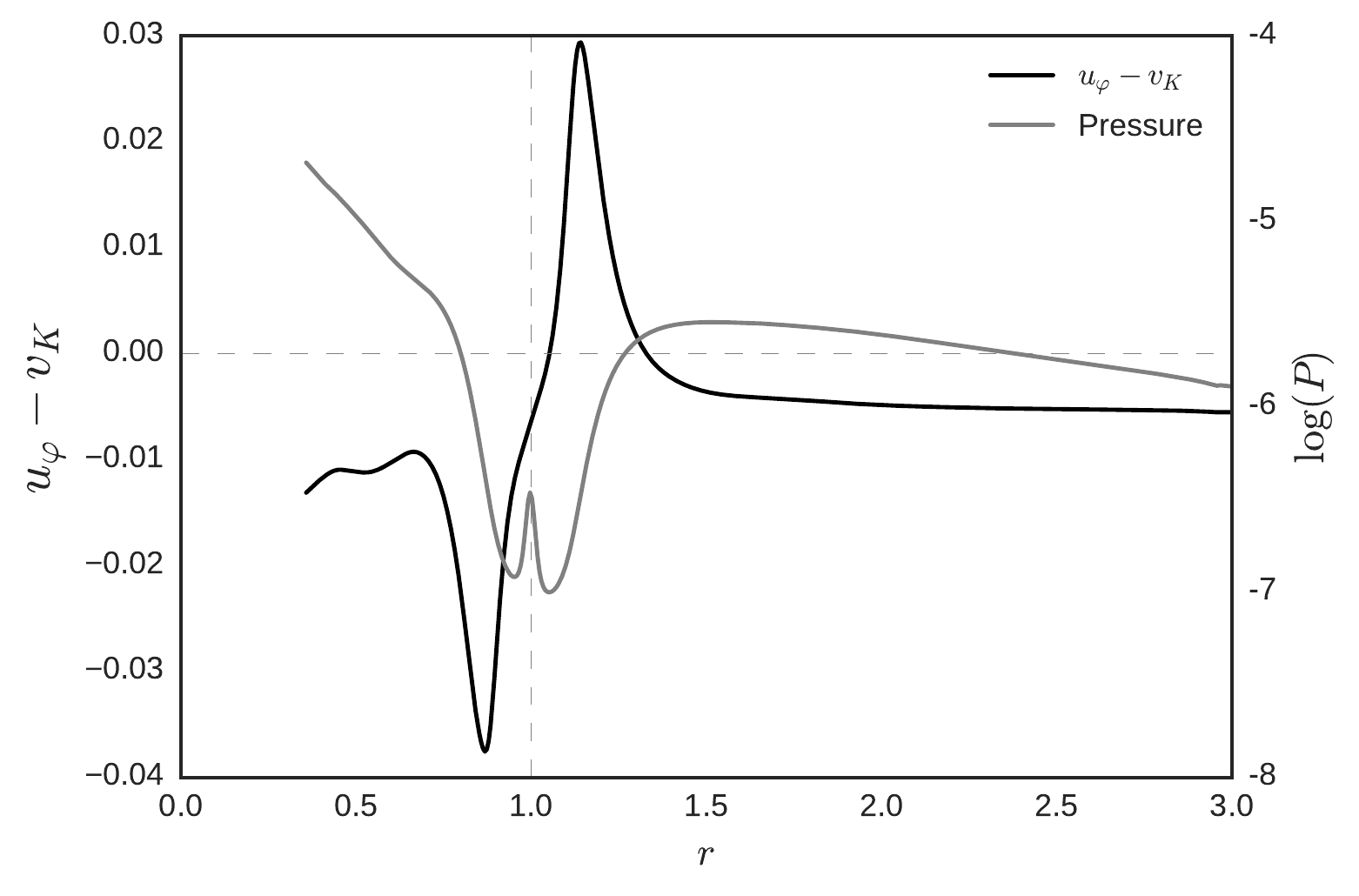}
\caption{Azimuthally averaged angular velocity (black line) and pressure (solid line) of the gas after evolving the disk for 10,000 orbits with a Jupiter-mass planet at $r=1$. The velocity is shown after subtraction of the Keplerian velocity $v_\mathrm{K}$ to highlight sub- and super-Keplerian rotation.}
        \label{fig:azivel}
\end{figure}

The evolved structure of the gas density after the first 10,000 orbits is shown in Figure~\ref{fig:2ddens} for the fiducial model, depicting the spiral wakes and the density depletion of around two orders of magnitude around the planet's orbit.
In Figure~\ref{fig:azivel}, we depict azimuthal averages of the gas rotation velocity in comparison to the Keplerian velocity (left axis), and the gas pressure profile (right axis). As expected from Equation~(\ref{eq:gasazi}), the azimuthal velocity of the gas becomes super-Keplerian at the outer edge of the gap, where the pressure gradient is positive.

In the following paragraphs, we study the transport of dust in the obtained environment, starting with a simplified, one-dimensional setup that is build upon our two-dimensional gas simulations. Afterwards, by comparison with two-dimensional \emph{dust} simulations, we illustrate the incompleteness of this treatment.

\subsection{One-dimensional studies} 
\label{subsec:1D}

\begin{figure}
\centering
\includegraphics[width=\columnwidth]{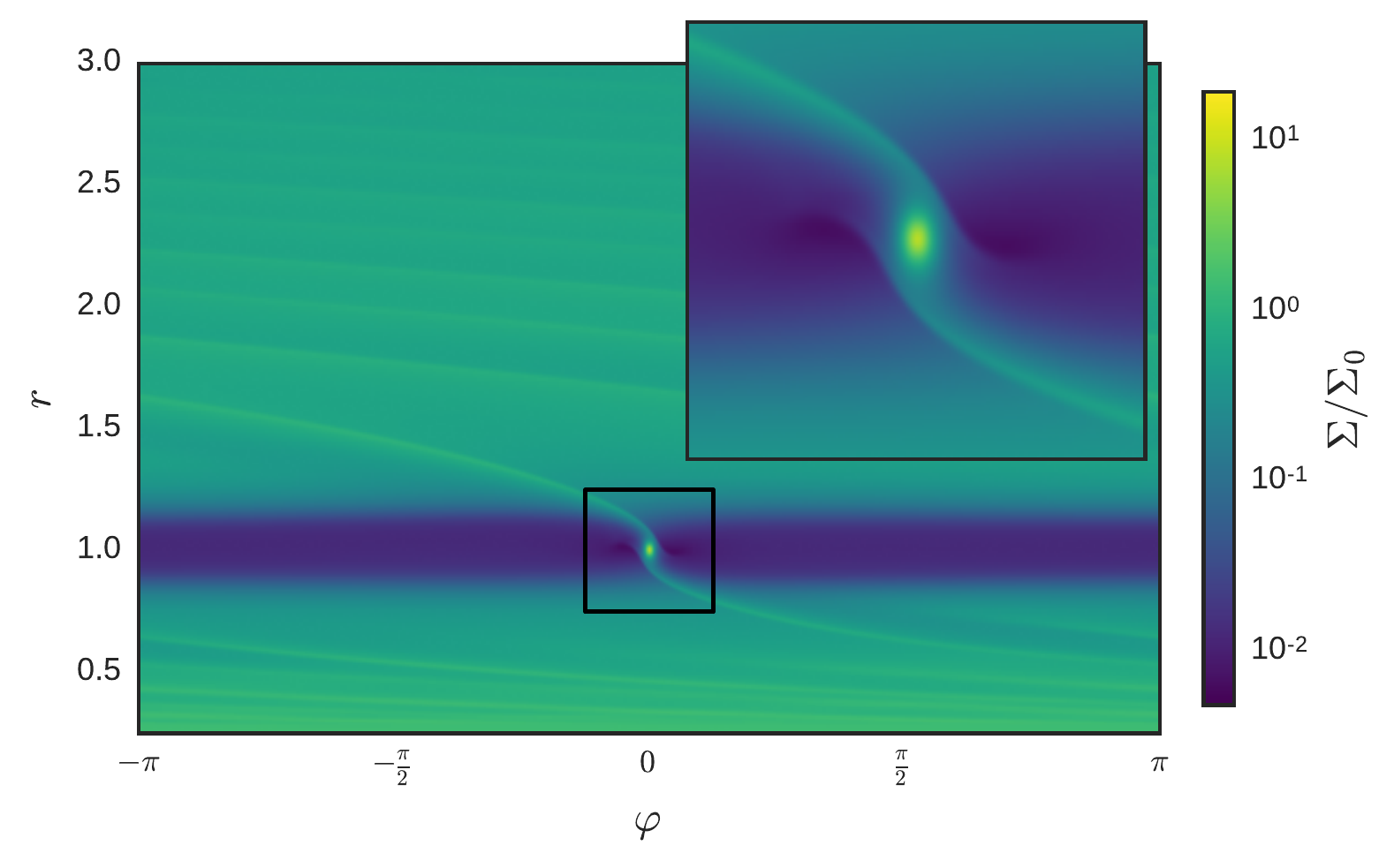}
\caption{Normalized gas surface density for a Jupiter-mass planet, that is, $q=M_\mathrm{P}/M_\star = 10^{-3}$. The distribution is shown after 10,000 orbits at the planet location.}
\label{fig:2ddens}
\end{figure}
Motivated by the one-dimensional description adopted in \citet{Rice2006}, and as a first approach, we study the filtration of dust in a simplified model that only retains the \emph{radial} variation of the steady-state disk structure. Figure~\ref{fig:2ddens} shows that, with the exception of the planet's location and the two spiral wakes, the gas density profile is mostly independent of azimuth, $\varphi$. Consequently, we take the azimuthal average of the angular velocity and surface density. Because of its highly asymmetric nature, we disregard a narrow angular sector (extending $\pm 5$ cells in azimuth) around the planet's location when averaging. We then impose a steady accretion flow of the gas (using the expression from Equation~(\ref{eq:vgr}) for the radial velocity), and furthermore remove the planet's potential from the disk since we are interested in exploring whether the dust filtering process can be characterized just in terms of planet-induced pressure perturbations.

In the next step, the dust is introduced in the outer disk, as specified in Equation~(\ref{eq:dustinit}), and while the gas profile stays unchanged as a static background, the dust evolves according to Equation~(\ref{eq:NS-dust}). We do not take diffusion into account, since the basic systematic trends are more easily discernible when neglecting the additional flux in Equation~(\ref{eq:contdust}).

\begin{figure}
\centering
\includegraphics[width=\columnwidth]{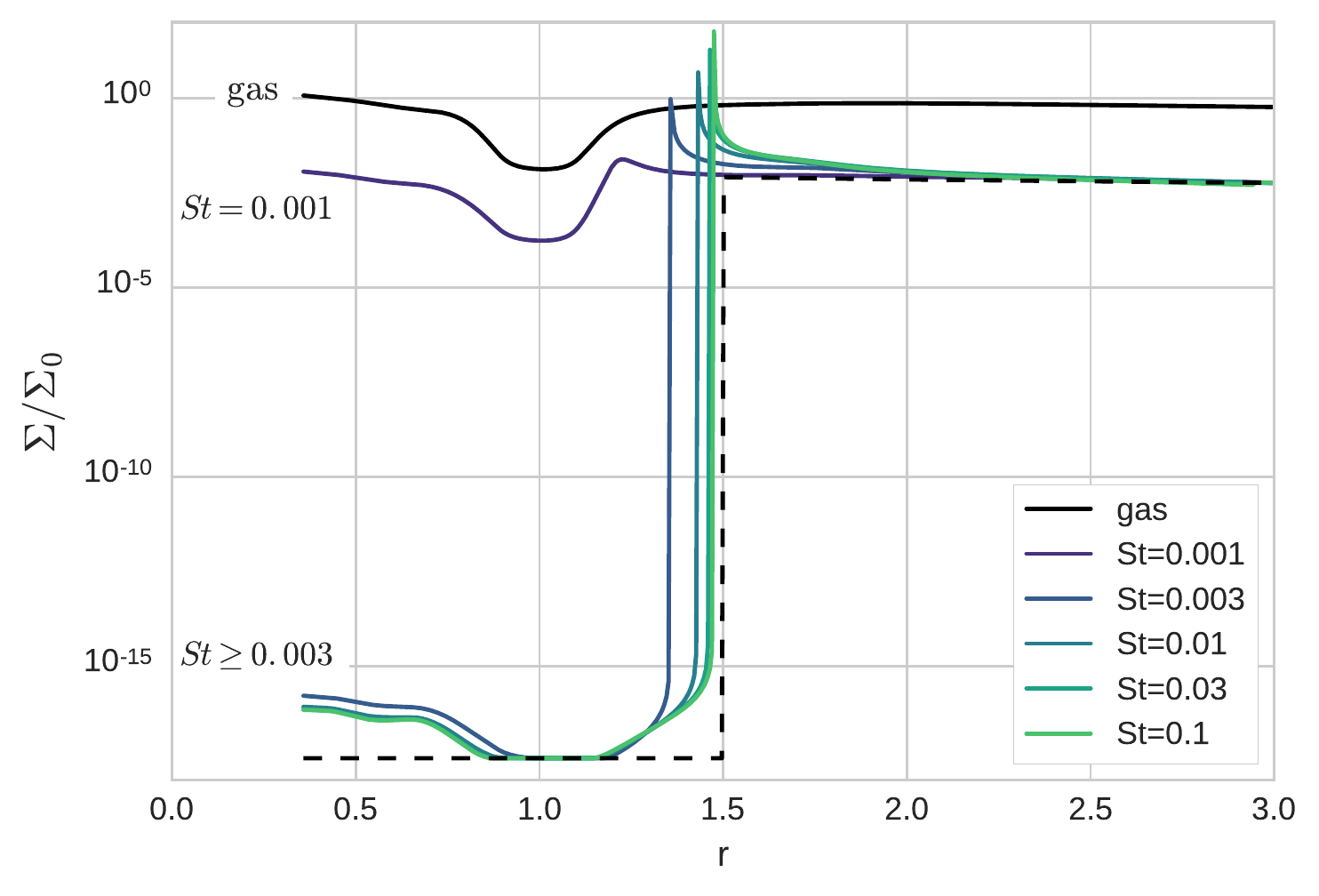}
\caption{1D simulation of the dust surface density evolution without diffusion. The dust is evolved for 10,000 orbits at $r_0$. The dashed line shows the initial distribution for the dust surface density of all species.}
\label{fig:1d}
\end{figure}

\begin{figure}
\centering
\includegraphics[width=\columnwidth]{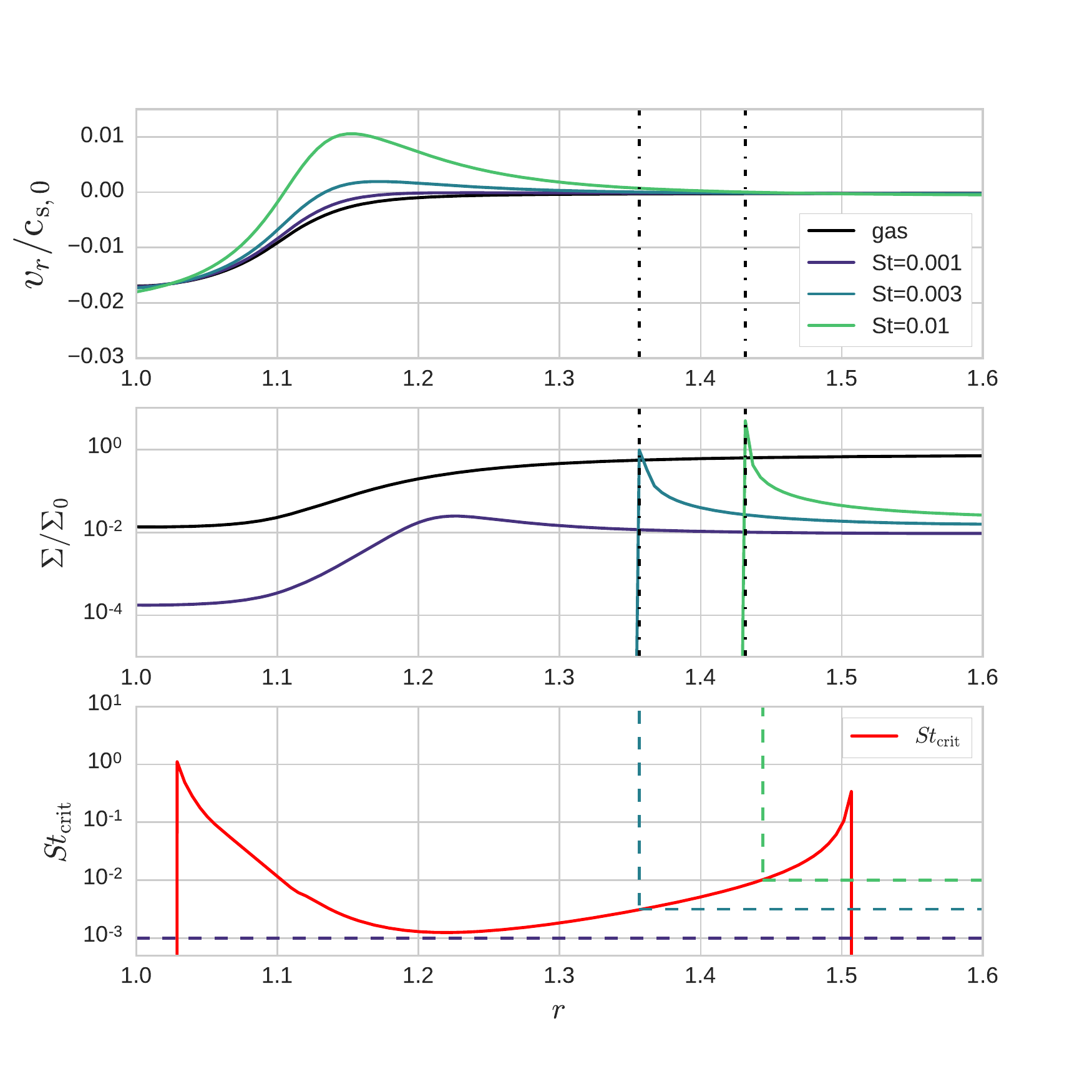}
\caption{Same simulation as in Figure \ref{fig:1d}. The top panel displays the radial velocity, the middle panel shows the dust surface density. The dashed dotted lines mark the location where the radial dust velocities turn positive for $\St=0.01$ (right) and $\St=0.003$ (left). The radial velocity of $\St=0.001$ is negative everywhere. The bottom panel shows the critical Stokes number calculated in Equation~(\ref{eq:Stcrit}), from which we can be estimate at which radial distance a certain dust species is stalled. The dashed lines are for direct comparison with the values used in the simulation.}
\label{fig:1d-zoom}
\end{figure}

Figure~\ref{fig:1d} illustrates the surface density profiles for several Stokes numbers after 10,000 orbits of evolving the dust in the stationary one-dimensional gas disk. The species with higher Stokes numbers drift faster towards the edge of the gap, but are stalled there. As there is no diffusion (and we do not include the back-reaction of the dust onto the gas), the dust can pile up infinitely. Therefore, in this setup the condition for transport through the gap simply is that $v_r$ remains negative for all radii. The upper two panels of Figure~\ref{fig:1d-zoom} show that the positions where the dust velocities have a zero-crossing coincide with the position where the density of the corresponding dust species has its cut-off. In contrast to this, the velocity of the grains with the smallest Stokes number considered here does not become positive anywhere and hence the dust is transported to the inner system. This shows that in 1D the outcome is entirely bimodal: Either, for small Stokes number, all the dust is allowed to pass and the dust-to-gas ratio in the inner system remains unaffected by the gap's existence -- or conversely for sufficiently large Stokes number, the grain population is completely filtered out.

Since, for every radial position in the disk, there exists a critical Stokes number, $\St_\mathrm{crit}$, for which one finds $v_r\left(r,\St_\mathrm{crit}\right)=0$, we can investigate where in the disk dust gets stalled depending on its Stokes number. From Equation~(\ref{eq:vdr}), we determine this threshold to be  \citep[cf.][]{Pinilla2012}
\begin{equation}
\St_\mathrm{crit}=u_r\left(\eta v_\mathrm{K}\right)^{-1}\,,
\label{eq:Stcrit}
\end{equation}
illustrated by the red line in the bottom panel of Figure~\ref{fig:1d-zoom}. We conclude that in 1D (and when neglecting dust diffusion), dedicated dust simulations are in fact unnecessary as the filtration behavior can easily be calculated from  the gas density and velocity structure itself. The aforementioned bimodal behavior is of course lost when diffusion is included. This case is discussed in detail in section~\ref{subsubsec:2Ddiffusion}, where we compare one- and two-dimensional models including dust diffusion.

\subsection{Two-dimensional simulations} 

As described in detail in Section~\ref{sec:condinit}, we have performed all simulations for 10,000 planet orbits with gas only. We now continue with presenting results from two-dimensional models, where we include typically five different dust species, and continue the evolution of the combined dust/gas system for a further 10,000 orbits.

\subsubsection{Fiducial Model} 
\label{subsec:fiducial}
\begin{figure*}
\centering
\includegraphics[width=.45\textwidth]{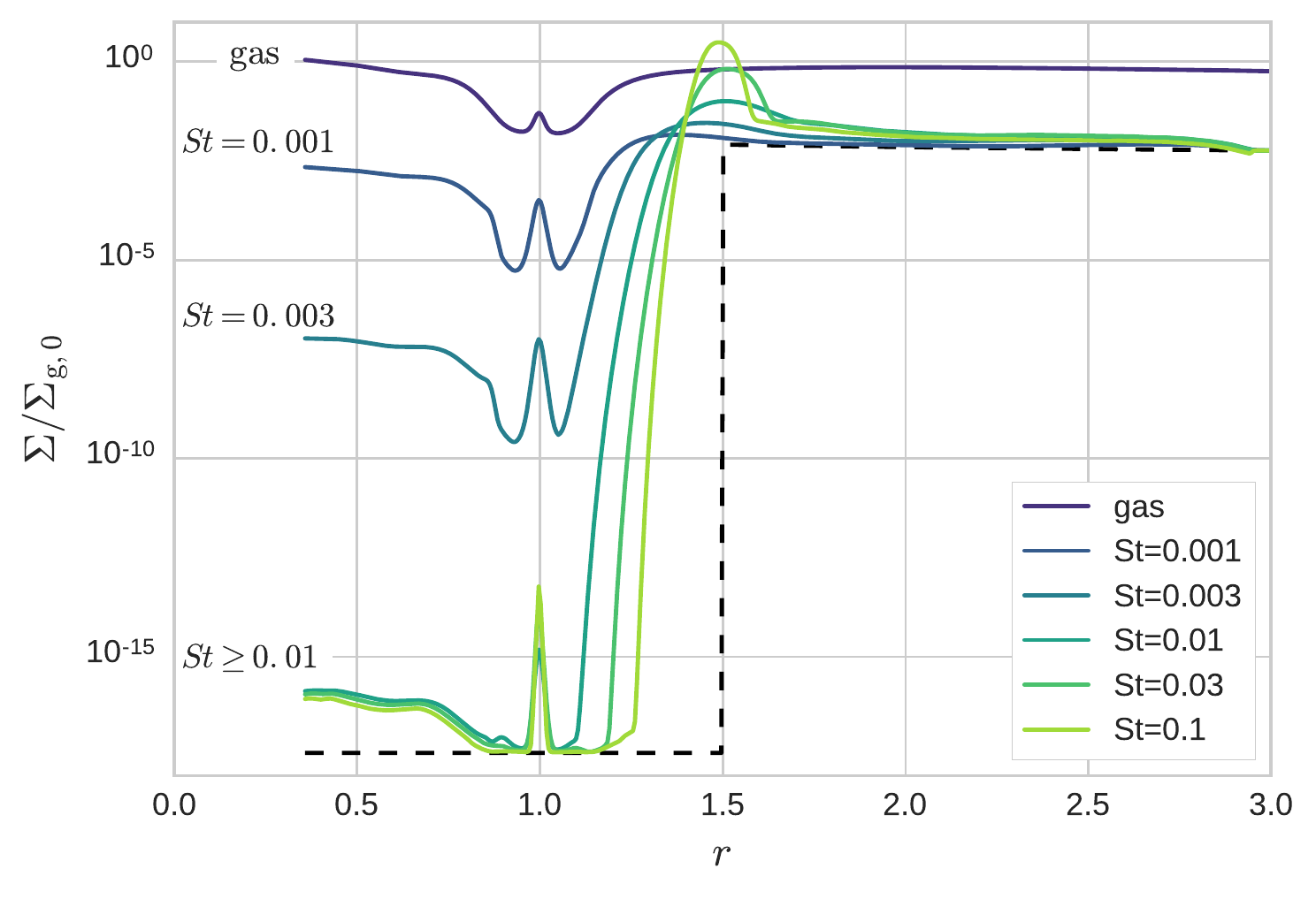}
\includegraphics[width=.45\textwidth]{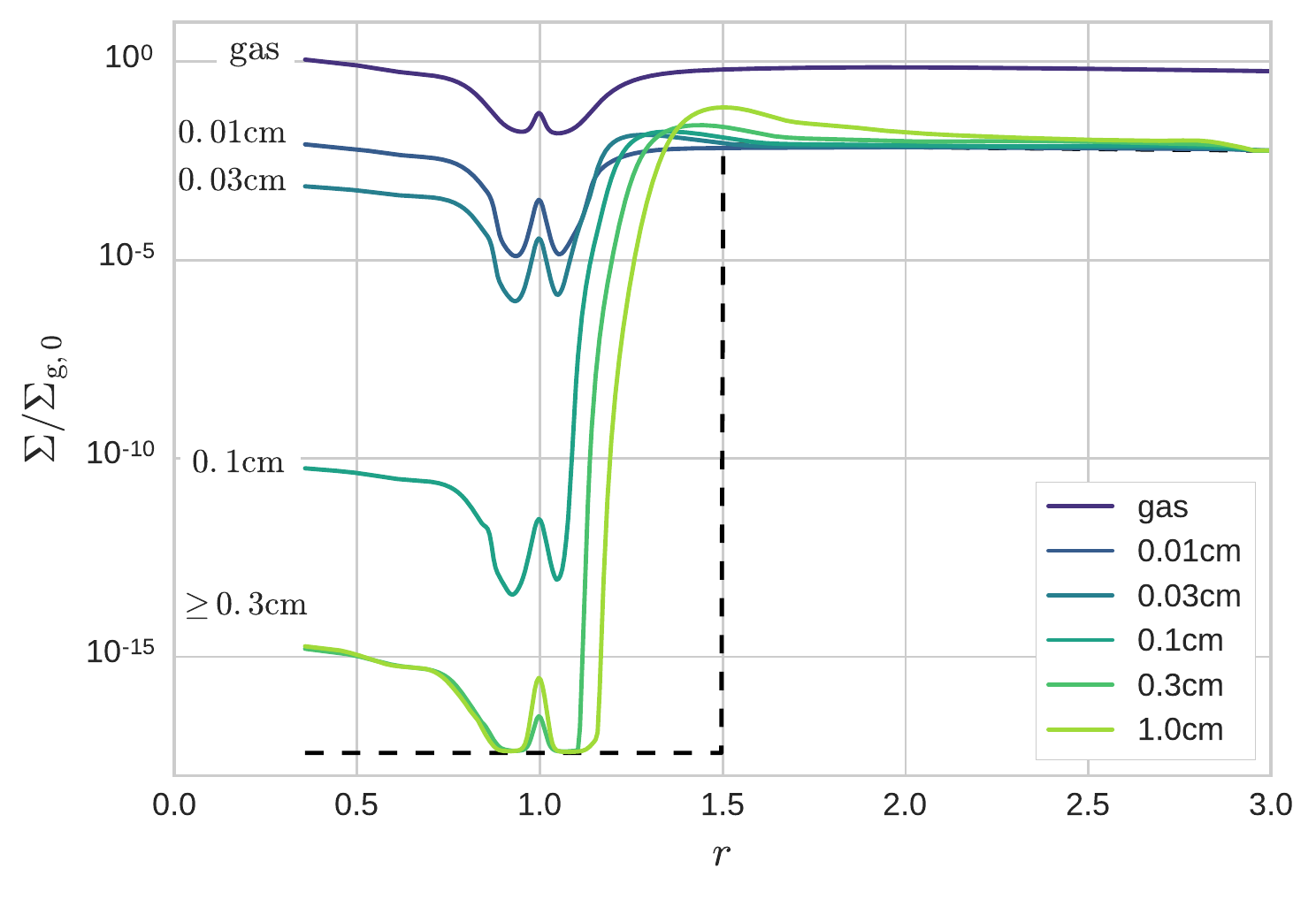}
\caption{Azimuthally averaged surface density profiles for gas and dust as a function of Stokes number (left panel) and grain size (right panel) for the fiducial model. The dashed line shows the distribution of the dust when it is added to the disk.}
\label{fig:2d-fiducial}
\end{figure*}

Our fiducial model represents a Jupiter-like planet in a modestly accreting disk with a typical radial surface density profile -- the detailed parameters are listed in Table~\ref{tab:para}. In the left panel of Figure (\ref{fig:2d-fiducial}) we plot the azimuthally averaged surface density distribution for different Stokes numbers after 10,000 orbits of combined evolution. The dust with $\St=0.01$ is effectively filtered out by a Jupiter-like planet, while smaller Stokes numbers are partially transported through the gap. This is in contrast to the 1D case, where the filtration was bimodal. Here, dust transport instead becomes \textit{gradually} more efficient when grains are more tightly coupled to the gas dynamics. Additionally, the density enhancement just outside of the gap (where large-enough dust grains become trapped) is smoother in the two-dimensional case. Both differences illustrate the fundamental shortcomings of the 1D model.

In order to make physically meaningful predictions, in what follows, we characterize the dust species in terms of their grain size, $a$, instead of their Stokes number, $\St$. Specifically, since $\St \propto a/\Sigma_\mathrm{g}$, dust grains of a constant size markedly change their interaction behavior with the gas when they enter the low-surface density gap region. Note that this behavior cannot be encapsulated in a fixed Stokes number. For the purpose of easier reference, we now specify our model by introducing concrete units. This allows us to label our results with a physically meaningful fiducial particle radius $a_{\rm fid}$, despite all simulations being performed in dimensionless units. With the Solar System in mind, we set the parameters of our model to values corresponding to a Jupiter-like planet around a solar mass star. In other words, we set $M_\ast = 1.0 \,\mathrm{M}_\odot$ and $r_\mathrm{P} = 5.2\, \textrm{AU}$. With this choice, the mass accretion rate of the fiducial model corresponds to $\dot{M} = 10^{-7}\,\mathrm{M}_\odot \, \mathrm{yr}^{-1}$. Note, however, that our results (such as the \emph{absolute} particle size) can be converted to apply to other PPD systems. We provide a simple scaling recipe for this in Appendix~\ref{ap:Appendix}.

The right panel of Figure~\ref{fig:2d-fiducial} shows the outcome in the fiducial model for a simulation with fixed particle sizes. The plot illustrates that dust grains bigger than about $a_\mathrm{crit}=0.3\,\mathrm{cm}$ do not replenish the inner parts of the disk at all, once a gap has been developed.

\begin{figure*}
\centering
\includegraphics[width=\textwidth]{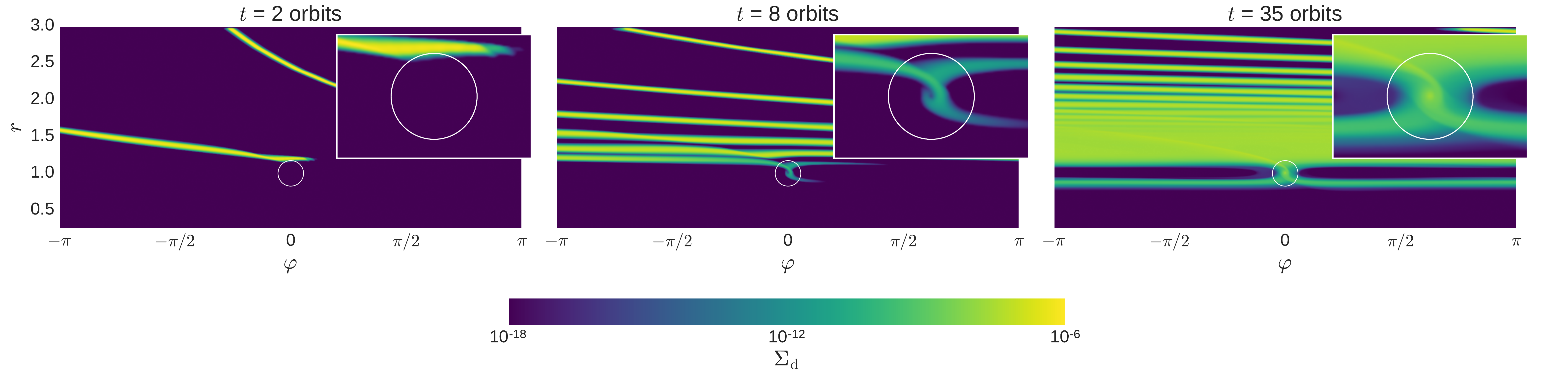}
\caption{Surface density of $a=10\,\mathrm{\mu m}$ dust that was initialized in only one azimuthal cell ($\varphi=3.0$) and for $r>1.3$. The panels show the distribution after 2, 8, and 35 orbits (at $r=1$), respectively. The frame is co-rotating with the planet, and the circle (with radius $2.5R_\mathrm{H}$) illustrates the planet's location.}
\label{fig:acc-process}
\end{figure*}

To demonstrate the importance of the two-dimensional flow around the planet, we set up a dust distribution that is initially confined to a narrow sector (spanning a single grid cell in azimuth), and covering the radial range outside the orbit of the planet. We introduce this configuration into the fiducial model for the gas after 10,000 orbits for which the surface density was shown in Figure \ref{fig:2ddens}. The described procedure allows us to directly follow the flow of the dust as it moves through the location of the planet.

Figure~\ref{fig:acc-process} shows three different snapshots during the evolution for dust grains of a size of $a=10\, \mathrm{\mu m}$ in a frame that is co-rotating with the planet (located at $r=1.0$ and $\varphi=0.0$). After two planet orbits, the initial profile is sheared out by the differential rotation. After about eight orbits, there is some dust injected into the inner system at the azimuthal location of the planet. After 35 orbits, the image already shows a considerable dust surface density inside of the planet's radial location.

To quantify the amount of mass that is transported to the inner system, we introduce a depletion parameter $\zeta$ that describes by what factor the dust density is decreased in the inner system in comparison to the case \emph{without} a planet. This is done by comparing the dust surface density, $\Sigma_\mathrm{d}$, in the presence of the planet at an inner location ($r_\mathrm{in}=0.4$) to the same quantity in the absence of the planet, here denoted by $\hat{\Sigma}_\mathrm{d}$. With this, the depletion factor as a function of grain radius becomes
\begin{equation}
\zeta(a) = \left. \frac{\Sigma_\mathrm{d}(a)}{\hat{\Sigma}_\mathrm{d}(a)}\right|_{r_{\rm in}}\,.
\label{eq:depletion}
\end{equation}
This will be the central quantity of interest when describing our results further, since it will allow a straightforward comparison between the different models.

\subsection{Influence of Model Parameters} 

In the previous section, we have discussed the fiducial model, for which the parameters are given in Table~\ref{tab:para}. Here, we address the question of how the outcome changes when a number of parameters are modified.

\begin{figure}
\centering
\includegraphics[width=.45\textwidth]{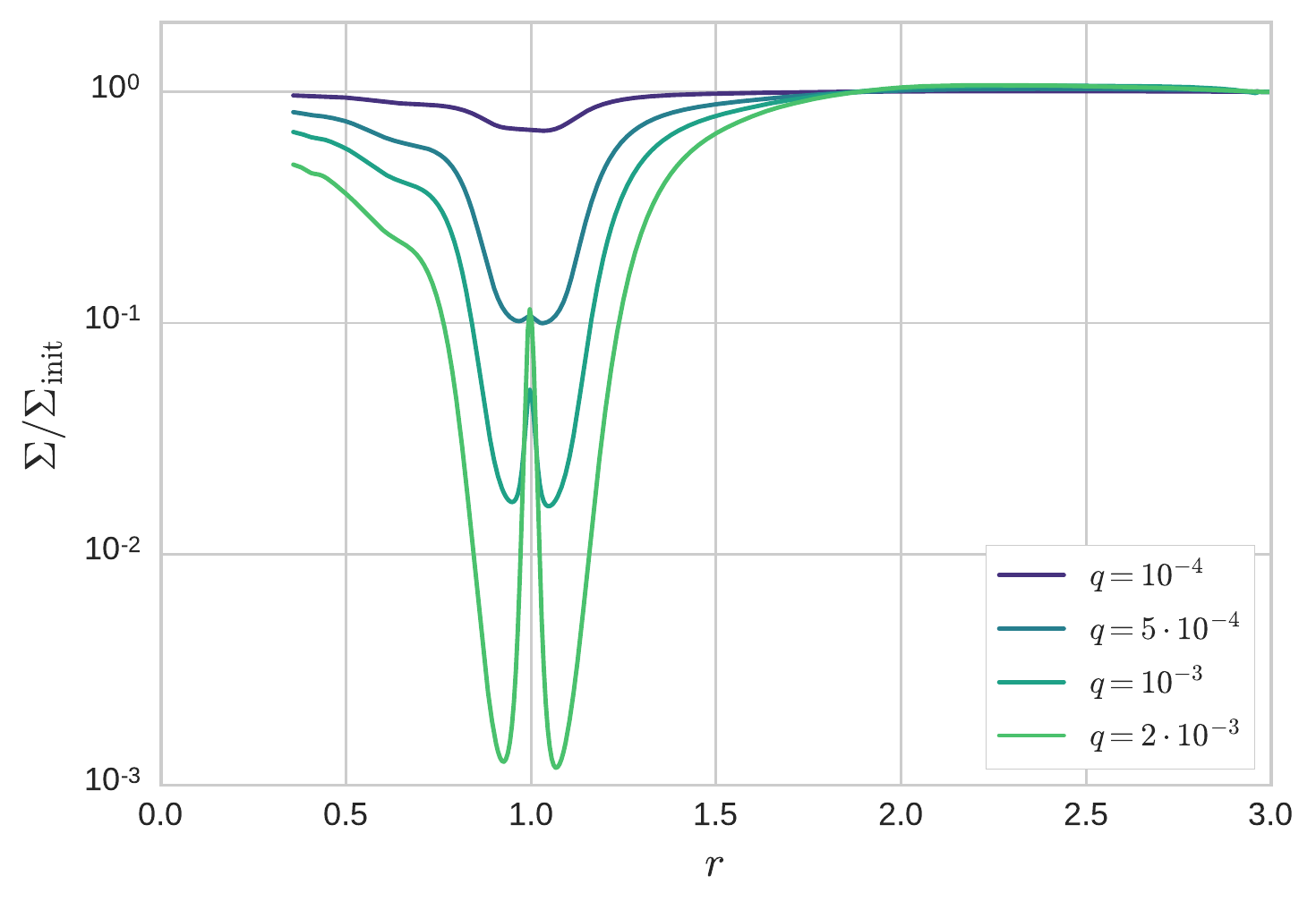}\\
\includegraphics[width=.45\textwidth]{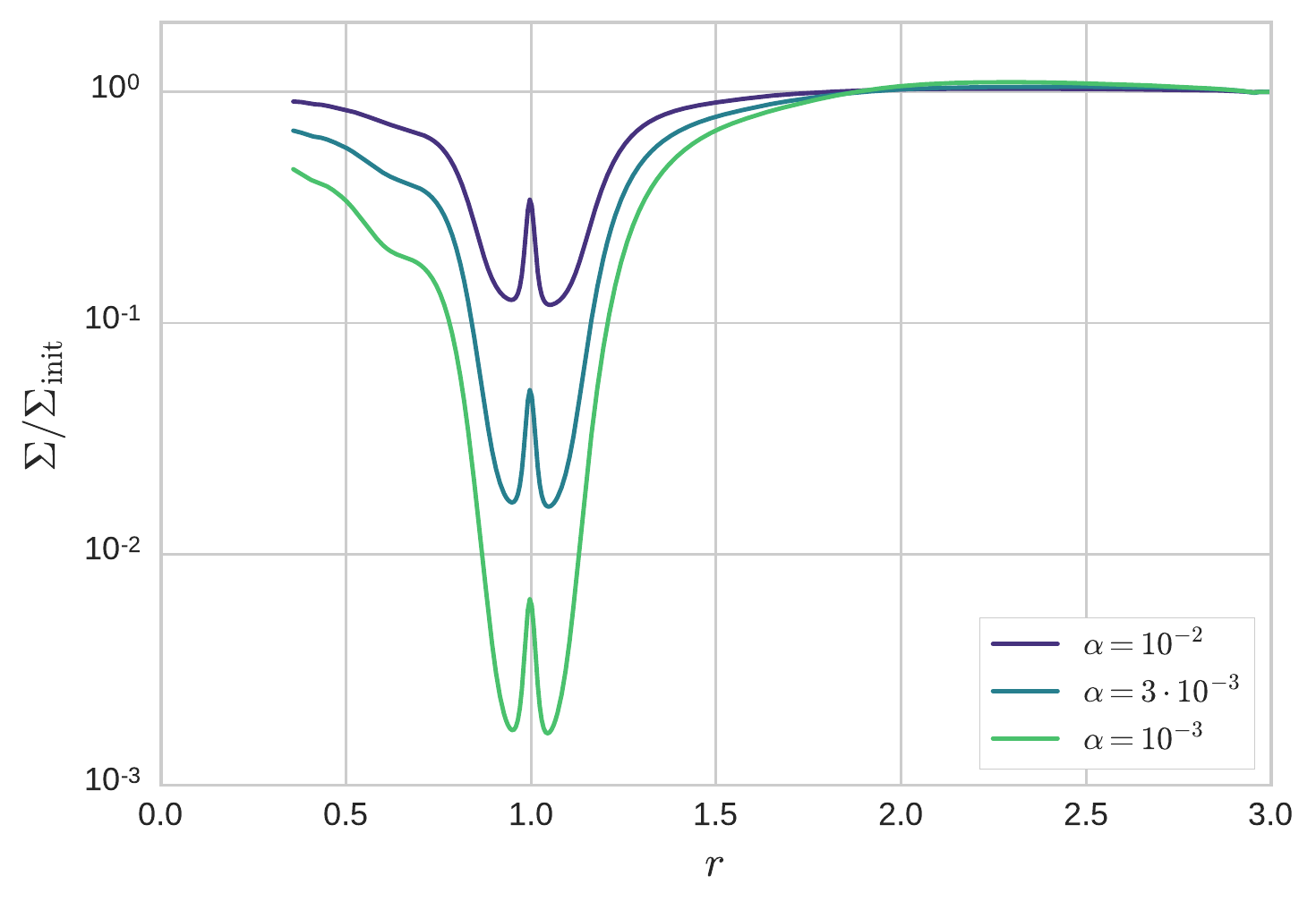}
\caption{Gas surface density for different planet masses (top) and $\alpha$-values (bottom) normalized by the initial profile without a planet. The gap structure is shown after 20,000 orbits.}
\label{fig:gasgap}
\end{figure}

We begin by highlighting that the effect of changing the gap structure is twofold: first of all, typically the deeper and narrower the gap is, the steeper is the pressure gradient at the outer gap edge, and with that $\eta\, v_\mathrm{K}$ increases in Equation~(\ref{eq:vdr}), making the dust filtration more efficient. Secondly, the deeper the gap, the higher becomes the Stokes number of a given species with fixed particle size when it is entering the gap. This means that even if a particle starts out tightly coupled to the gas outside of the gap region, it can decouple simply by the decreasing gas density around it when coming closer to the planet's orbit. Consequently, a particle that may be in the low-Stokes number regime in the outer regions of the disk, can still be subject to filtration when it reaches the gap.

With regard to the overall structure of the gas density in the gap region, \cite{Crida2006} give an analytical criterion for a planet to be able to open a gap in the gas, that is
\begin{equation}
\frac{3}{4}\frac{H}{R_\mathrm{H}} + \frac{50 \nu}{q r_\mathrm{P}^2 \Omega_\mathrm{P}} \lesssim 1\,,
\end{equation}
where $R_\mathrm{H}$ is the planet's Hill radius. Within the $\alpha$-viscosity framework, free parameters occurring in this formula can be reduced to $q$, $\alpha$ and $H$. Several studies of the gap structure \citep[e.g.][]{Fung2014,Kanagawa2016} have shown that, in steady-state, the depletion of the gas surface density generally increases for higher planet mass ratios, $q$, while increasing the viscosity parameter, $\alpha$, or the scale height, $H$, leads to a shallower gap. The equation above predicts that, for our fiducial model, a full gap is opened in the gas. In Figure~\ref{fig:gasgap}, we display that trend by showing the azimuthally averaged gas structure that we obtain after 20,000 orbits (at the planet location) for different parameters $q$ (top panel) and $\alpha$ (bottom panel).

\subsubsection{Dependence on planet mass}\label{sec:planetmass} 

\begin{figure*}
\centering
\includegraphics[width=.45\textwidth]{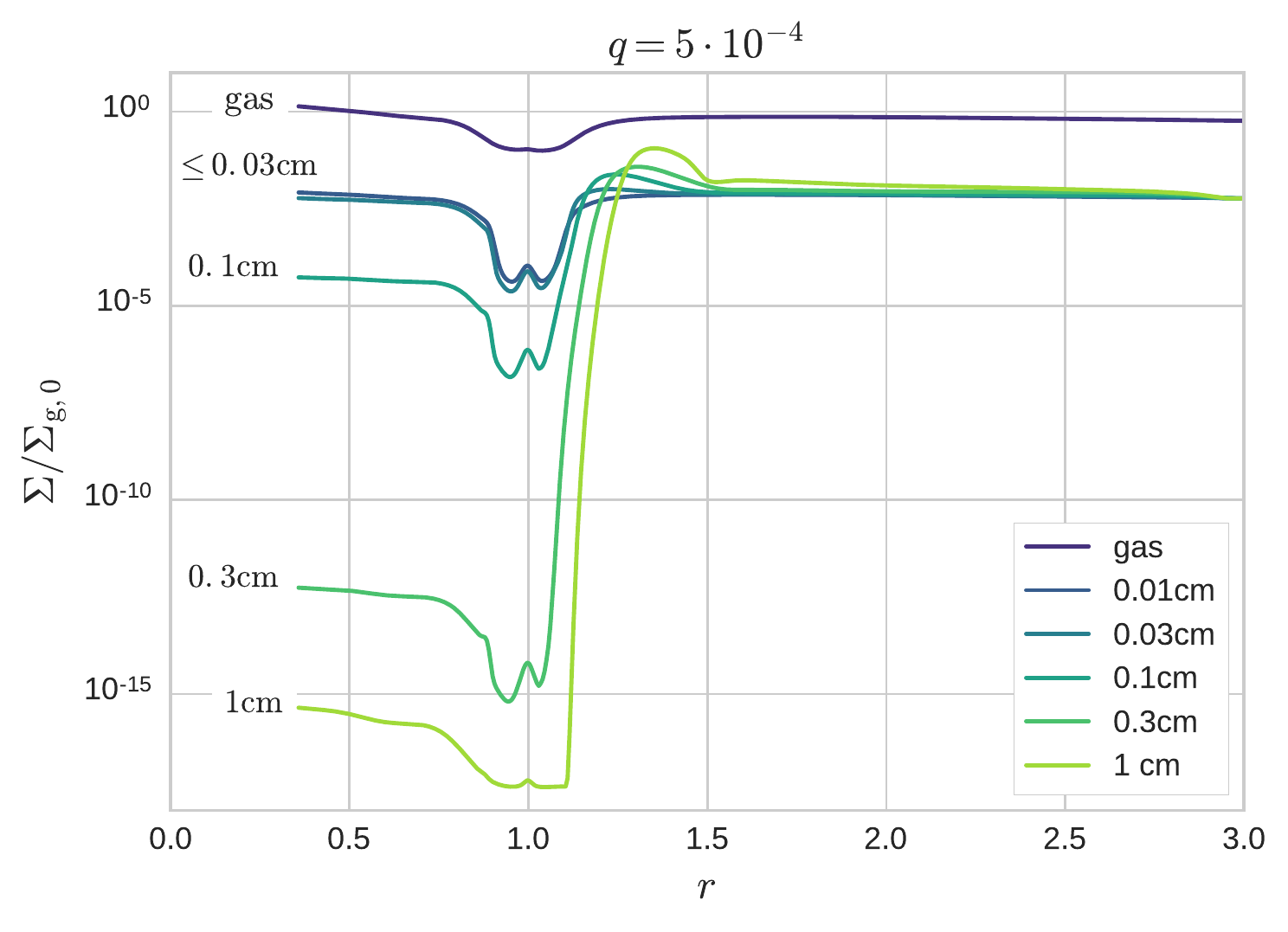}
\includegraphics[width=.45\textwidth]{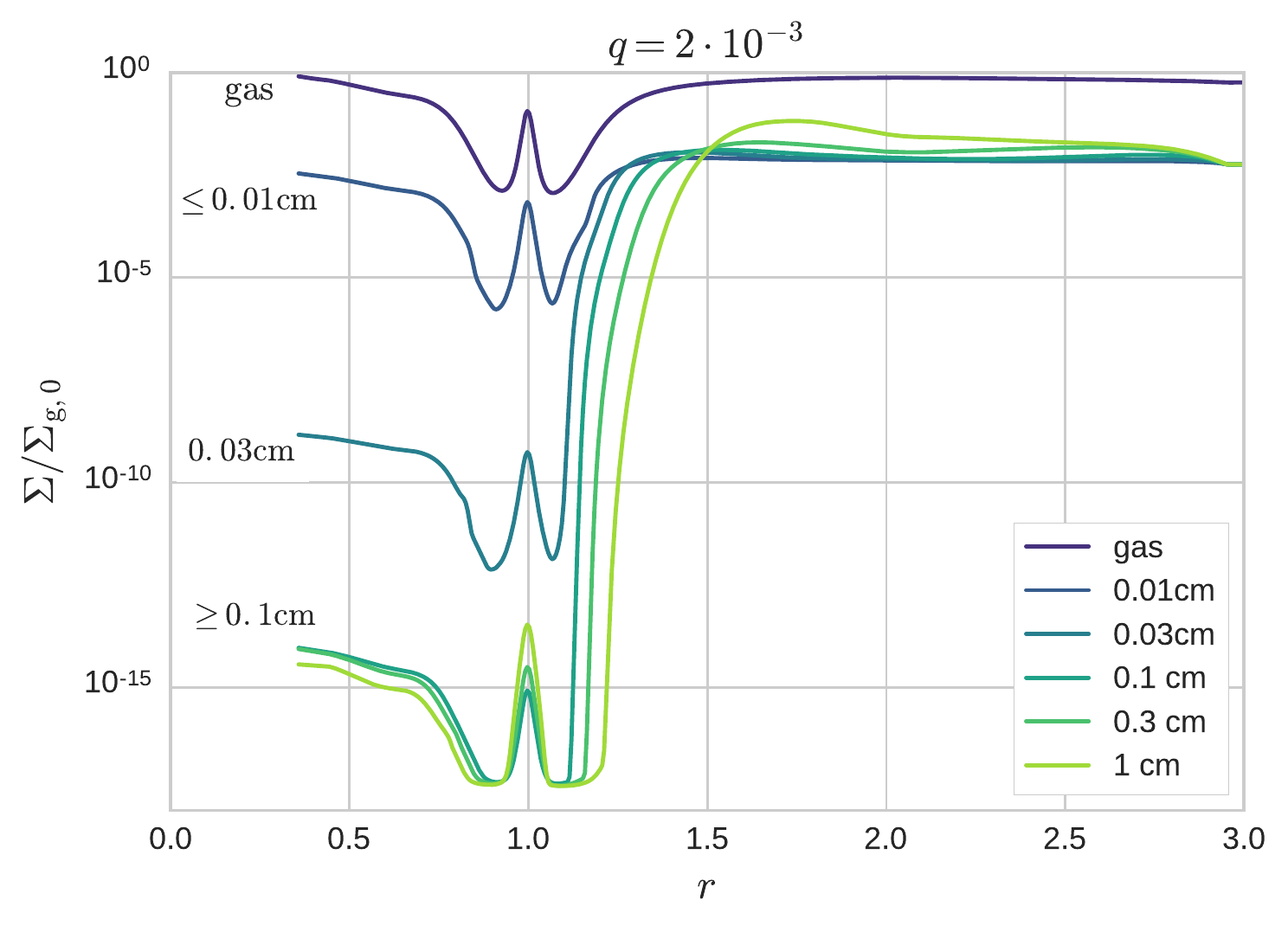}
\caption{Azimuthally averaged surface density profiles of gas and dust after 20,000 orbits for $q=5\times10^{-4}$ (left) and $q=2\times10^{-3}$ (right).}
\label{fig:q}
\end{figure*}

With the gap in the gas becoming more depleted for higher planet masses, we expect the filter to be more efficient for more massive planets. In the left and right panel of Figure~\ref{fig:q}, we show the azimuthally averaged surface density distribution for the case of $q=5\times10^{-4}$ and $q=2\times 10^{-3}$, respectively. Note that the range of grain sizes differs in the two plots. The figures show an important behavior: The density of a dust species that is filtered out by the gap does not necessarily pile up steeply at the location of the pressure bump. We can see this in the right panel of Figure~\ref{fig:q}, where the planet mass is high enough to deplete the surface density by almost three orders of magnitude compared to the region of the pressure trap. This means that also the Stokes number of a dust grain of fixed size is increasing by three orders of magnitude. Therefore, it is possible in the region where the trapping takes place (that is, just outside the gap) that a dust grain has a small enough Stokes number to become tightly coupled to the gas -- as a consequence, it does not get trapped. When approaching the planet's orbit, however, its Stokes number increases rapidly, decoupling it from the viscous accretion flow.

In the left panel of Figure~\ref{fig:q}, we see that the grains that get filtered out also pile up at the outer edge of the gap. The explanation is the same as before, only now the depletion in the gap is merely one order of magnitude, so that the change in Stokes number is not sufficiently high to filter out grains that are not piling up.

\begin{figure}
\centering
\includegraphics[scale=0.55]{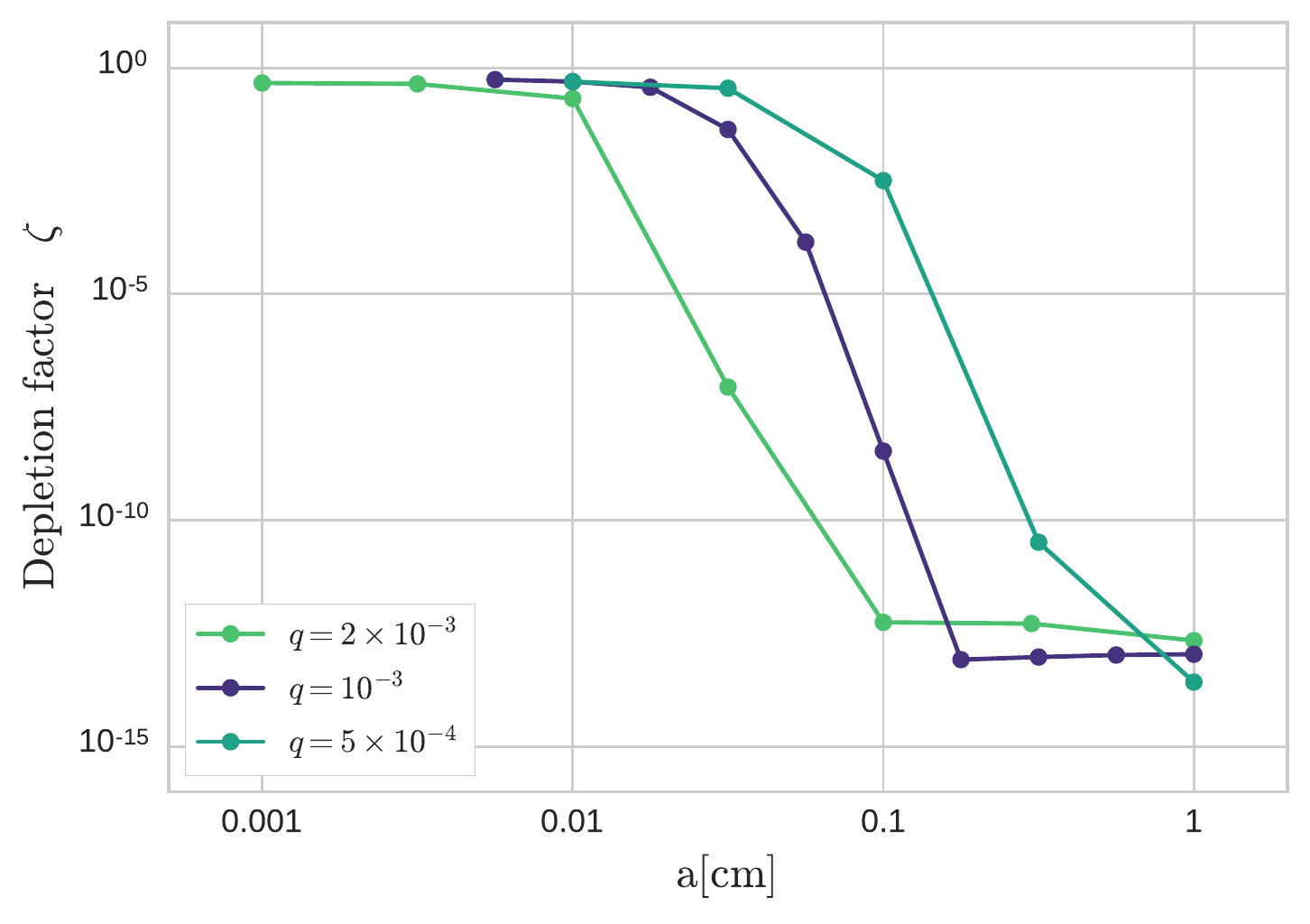}\vspace{-1ex}
\caption{Effective dust permeability of the gap as a function of grain size and for different values of the relative planet mass, $q$. The depletion factor is evaluated after the dust was evolved for 10,000 planet orbits.}
\label{fig:qfilter}
\end{figure}

To compare the size dependency of the filtration for different planet masses, we take a look at the size-dependent depletion factor (defined in Equation~(\ref{eq:depletion}) above), for different values of their dimensionless mass, $q$. Accordingly, Figure~\ref{fig:qfilter} shows that while the shape of the function remains about the same, it is shifted towards smaller grain sizes for higher planet masses. This trend can easily be understood by the reduction of the effective Stokes number, of particles of a fixed physical size, when entering the low-density gap region, which is deeper for a more massive planet, and the steeper pressure gradient in that case.

\subsubsection{Dependence on disk viscosity} 

\begin{figure*}
\centering
\includegraphics[width=.45\textwidth]{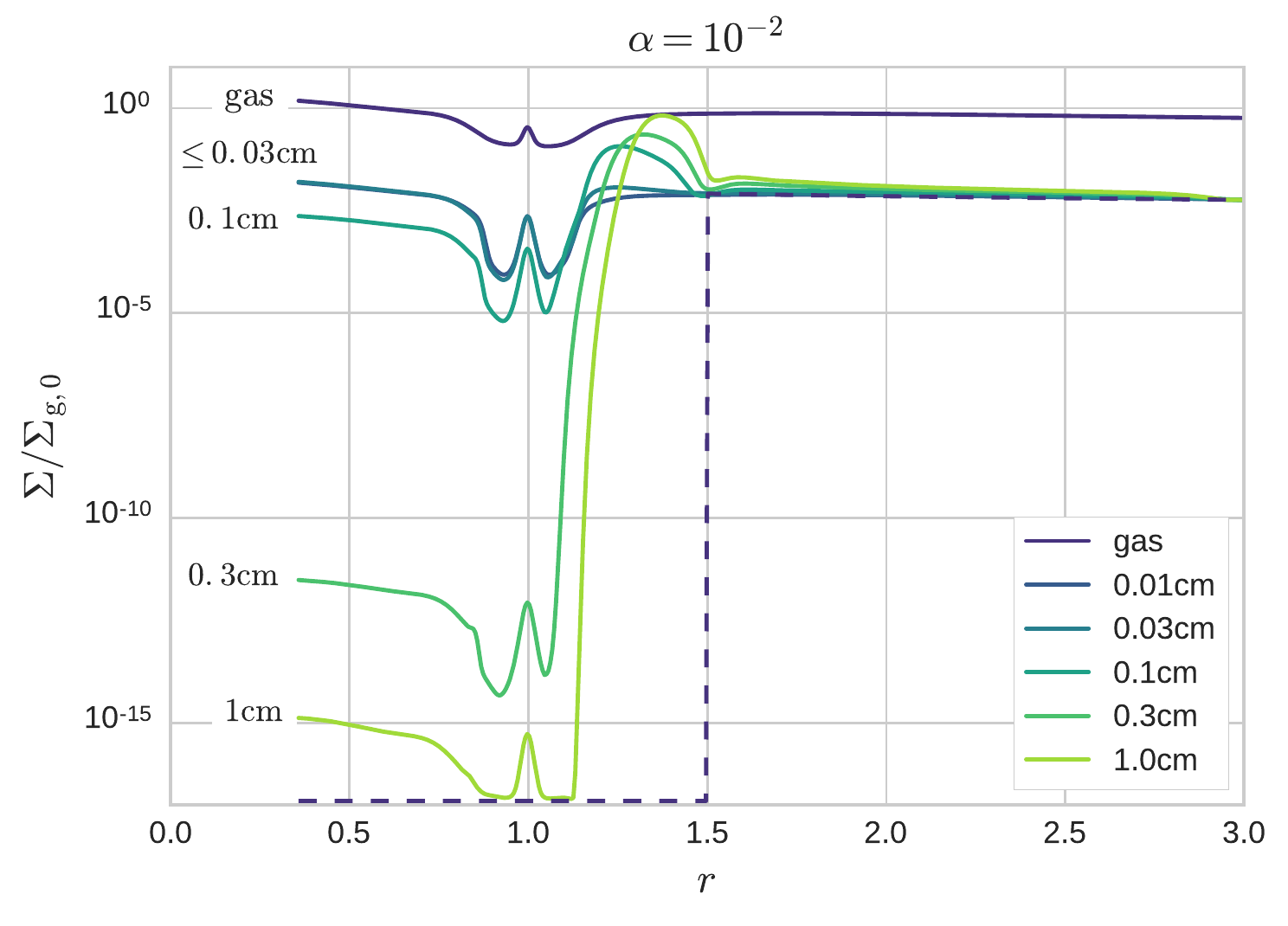}
\includegraphics[width=.45\textwidth]{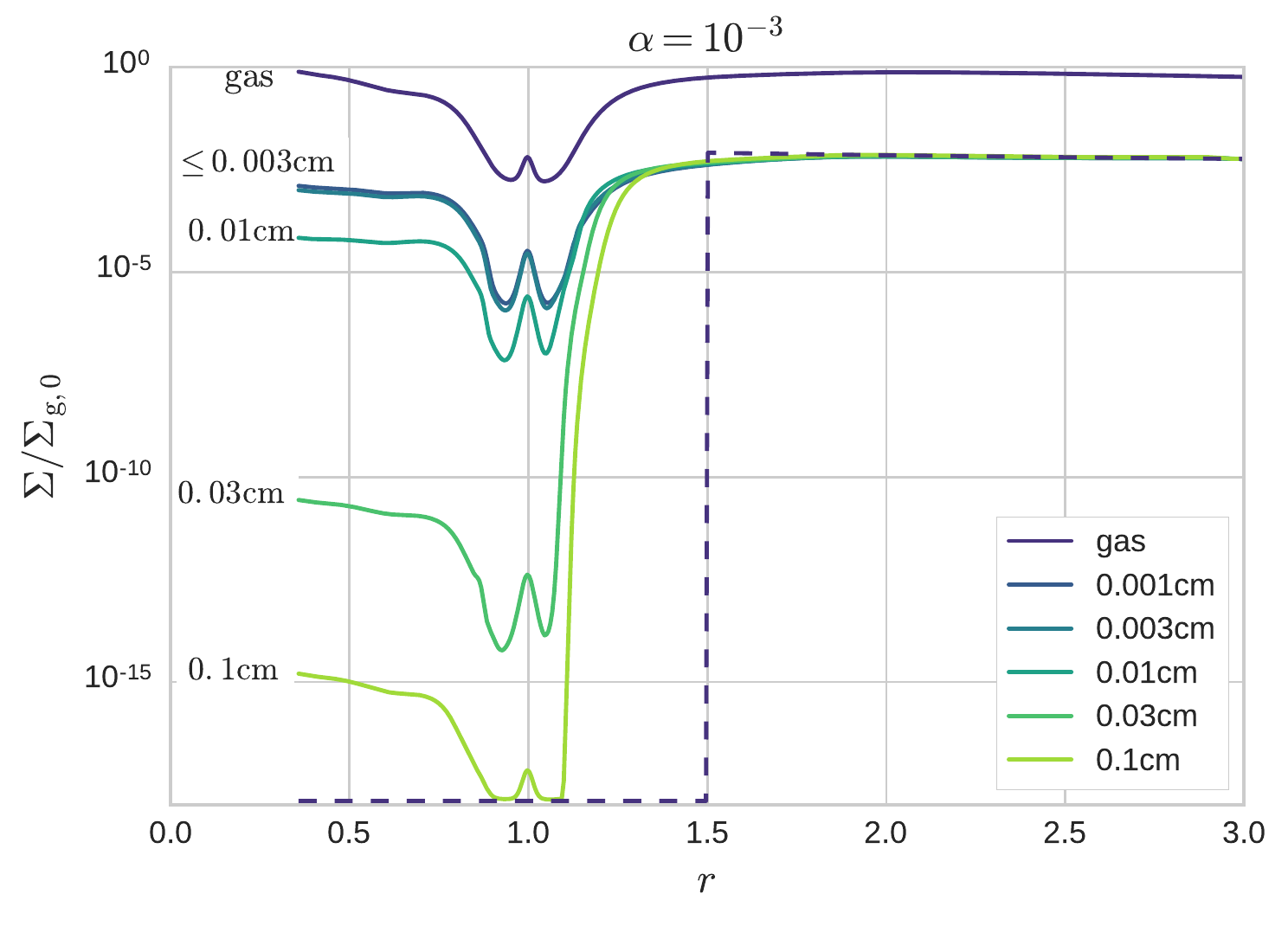}
\caption{Azimuthally averaged surface density profiles of gas and dust after 20,000 orbits for $\alpha=10^{-2}$ (left) and $\alpha=10^{-3}$ (right). The dashed line shows the initial distribution of the dust species when they were introduced after 10,000 orbits.}
\label{fig:alpha}
\end{figure*}

Changing the viscosity of the disk has multiple effects on the transport of the dust. First of all, as can be seen in the lower panel of Figure~\ref{fig:gasgap} above, the amount of viscosity changes the structure of the gas gap, such as, the depth and width of the gap. Secondly, the radial gas velocity as given by Equation~(\ref{eq:vgr}) and the surface density of the gas as given by Equation~(\ref{eq:sigma-g}) are modified. The local disk structure is important when considering the dust transport. A deeper gap will produce a stronger filter for dust, a smaller radial velocity will slow down the transport even in the unperturbed regions, and a higher gas density affects the Stokes number associated with particles of a given size.

Figure~\ref{fig:alpha} shows how modifying the viscosity parameter by a factor of three up and down (that is, to $\alpha=10^{-2}$ and $\alpha=10^{-3}$, respectively) changes the resulting dust distribution. Note that the range of sizes depicted in each of the figures again is different by an order of magnitude between the two plots. As described in detail in Section~\ref{sec:planetmass}, we again see that for $\alpha=10^{-3}$ grains that are filtered out do not necessarily pile up around a narrow annulus, as was the case in the 1D model -- the explanation is the same as in the previous section.

\begin{figure}
\centering
\includegraphics[scale=0.55]{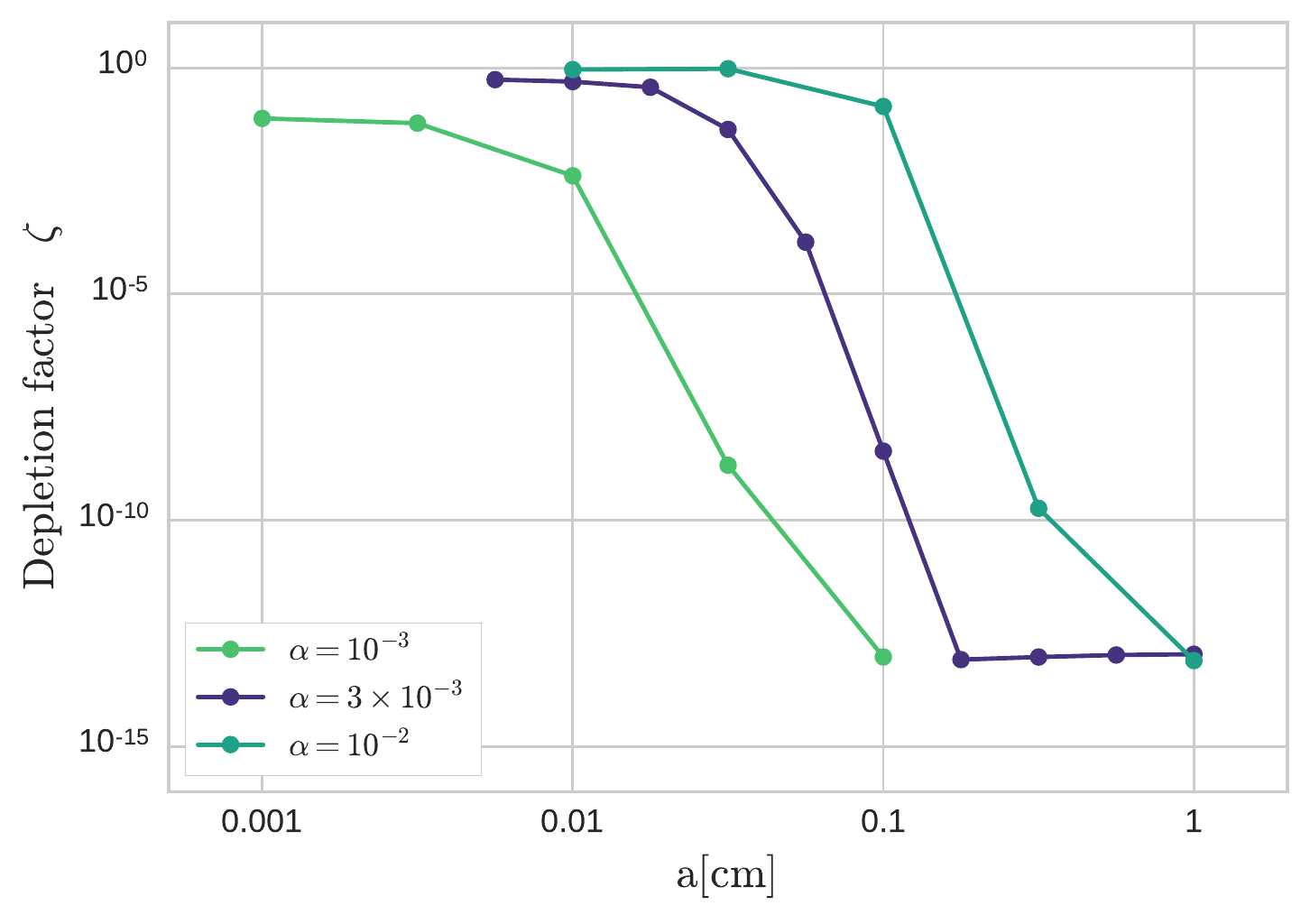}\vspace{-1ex}
\caption{Effective dust permeability of the gap as a function of grain size and for different values of the turbulence parameter, $\alpha$.}
\label{fig:alphafilter}
\end{figure}

As before, we consider the abundances in the inner system at late times and compare the depletion for different parameter values, as is shown in Figure~\ref{fig:alphafilter}. It is noteworthy that for the low viscosity case ($\alpha=10^{-3}$), the surface density distribution of the smallest dust is not yet in equilibrium. This is because these dust grains are very tightly coupled to the gas and from Equation~(\ref{eq:timescale}), one finds that the viscous timescale in the unperturbed case is about $40,000$ orbits for this value of $\alpha$, which is not covered by our simulation yet.

\subsection{The effect of dust diffusion}\label{subsubsec:2Ddiffusion} 

Until now, we have neglected the diffusive dust flux, denoted by $\mathbf{j}$ in Equation~(\ref{eq:contdust}). In this section, we take a closer look at the effect of particle diffusion on the permeability of the planetary gap. In the following, we hence set the diffusive flux to the expression given by Equation~(\ref{eq:diffflux}).

\begin{figure*}
\centering
\includegraphics[height=.333\textwidth]{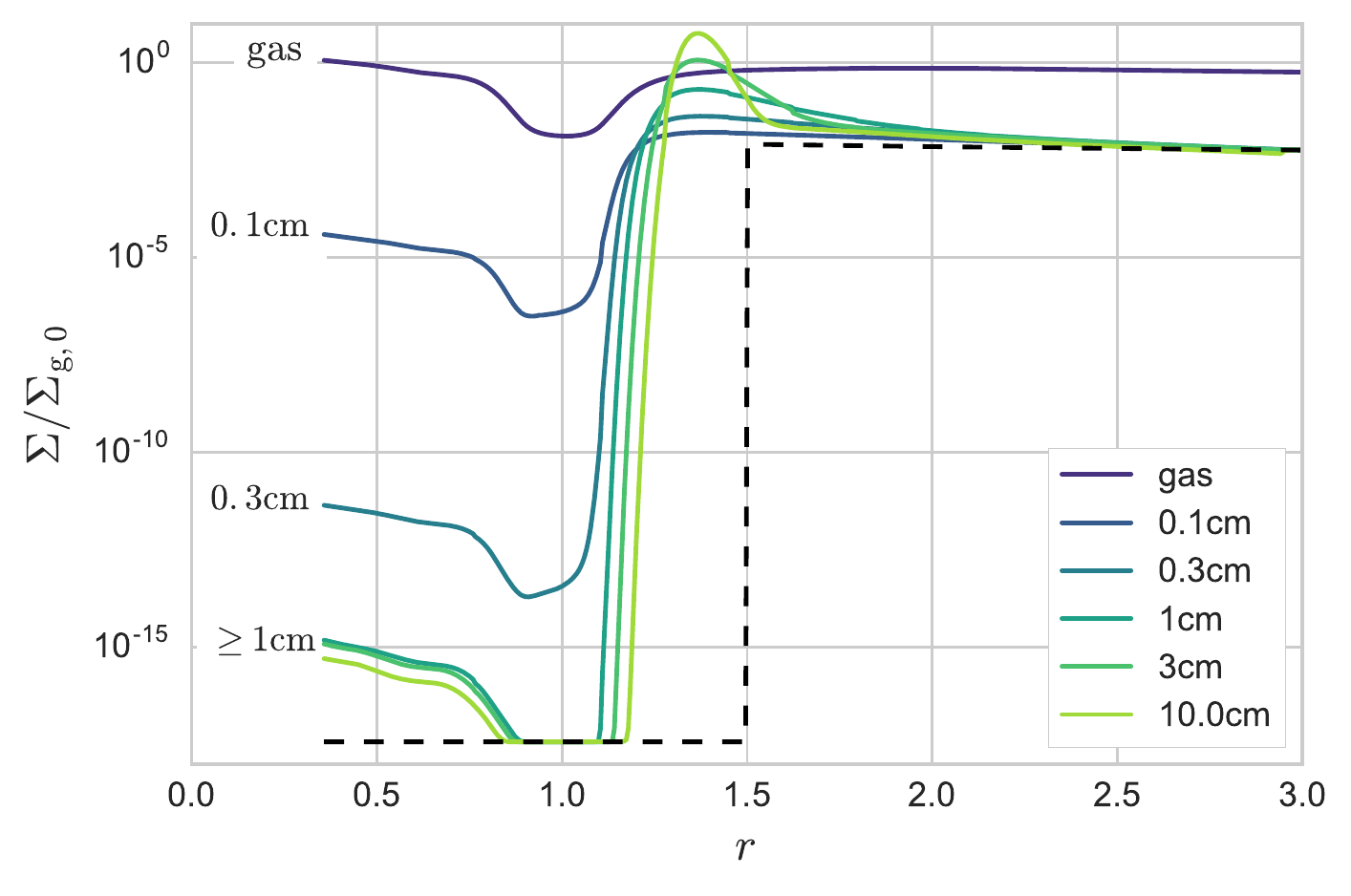}
\includegraphics[height=.333\textwidth]{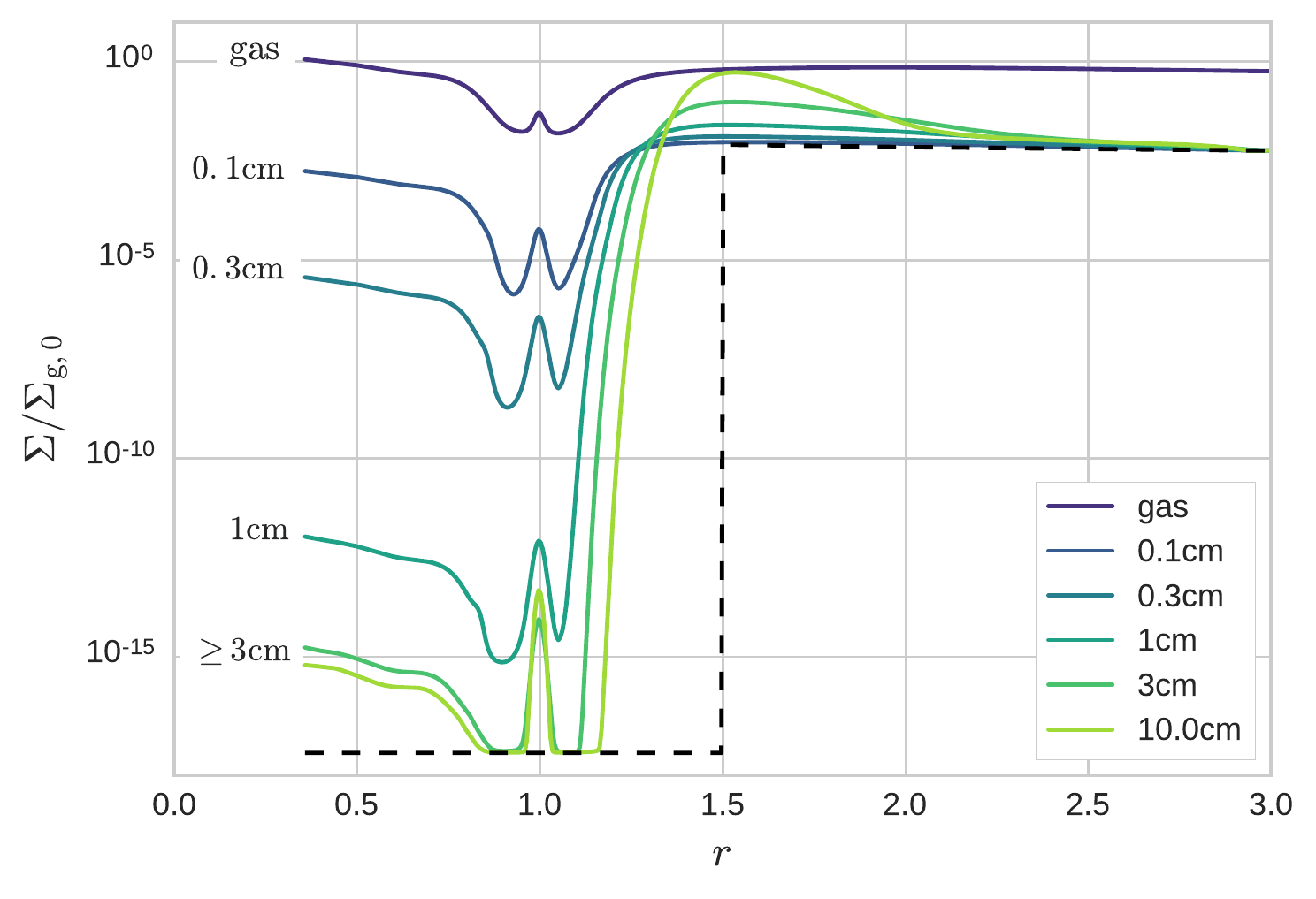}
\caption{Azimuthally averaged surface density profiles of gas and dust after 20,000 orbits for the fiducial model, but now \emph{with} particle diffusion enabled. The left and right panels show the result for a 1D and 2D simulation, respectively.}
\label{fig:sizediff}
\end{figure*}

The right-hand panel of Figure~\ref{fig:sizediff} shows the resulting density distributions in the case accounting for dust diffusion (with a Schmidt number of unity) for our fiducial two-dimensional model and for fixed particle sizes. Since at the outer gap edge the surface density distribution of the dust becomes very steep the effect of the diffusion is strongest there. This causes an additional flux of dust into the gap and causes the peaks of the distributions to become smoother. This effect is less pronounced in the simplified one-dimensional case\footnote{see Section~\ref{subsec:1D} for details on how these runs were constructed}, shown in the corresponding left-hand panel of Fig.~\ref{fig:sizediff}, and where the pile-up outside of the gap edge is more markedly seen. A potential explanation for why the dust peak in 1D is narrower than in 2D may be offered by the following reasoning: as argued in Section~\ref{subsec:1D}, the pile-up outside the gap is obtained at the locus where the effective radial velocity of the dust fluid becomes zero, and the peak is subsequently smoothed-out by the dust diffusion. While the latter effect is the same in both 1D and 2D, the difference may stem from that, in 2D, the $v_r(r,\phi)=0$ locus is not necessarily at a fixed radius, but may be a function of the coordinate $\phi$. In this case, the bump will be blurred-out in radius when performing the azimuthal average, leading to an apparent broader profile.

\begin{figure}
\centering
\includegraphics[scale=0.55]{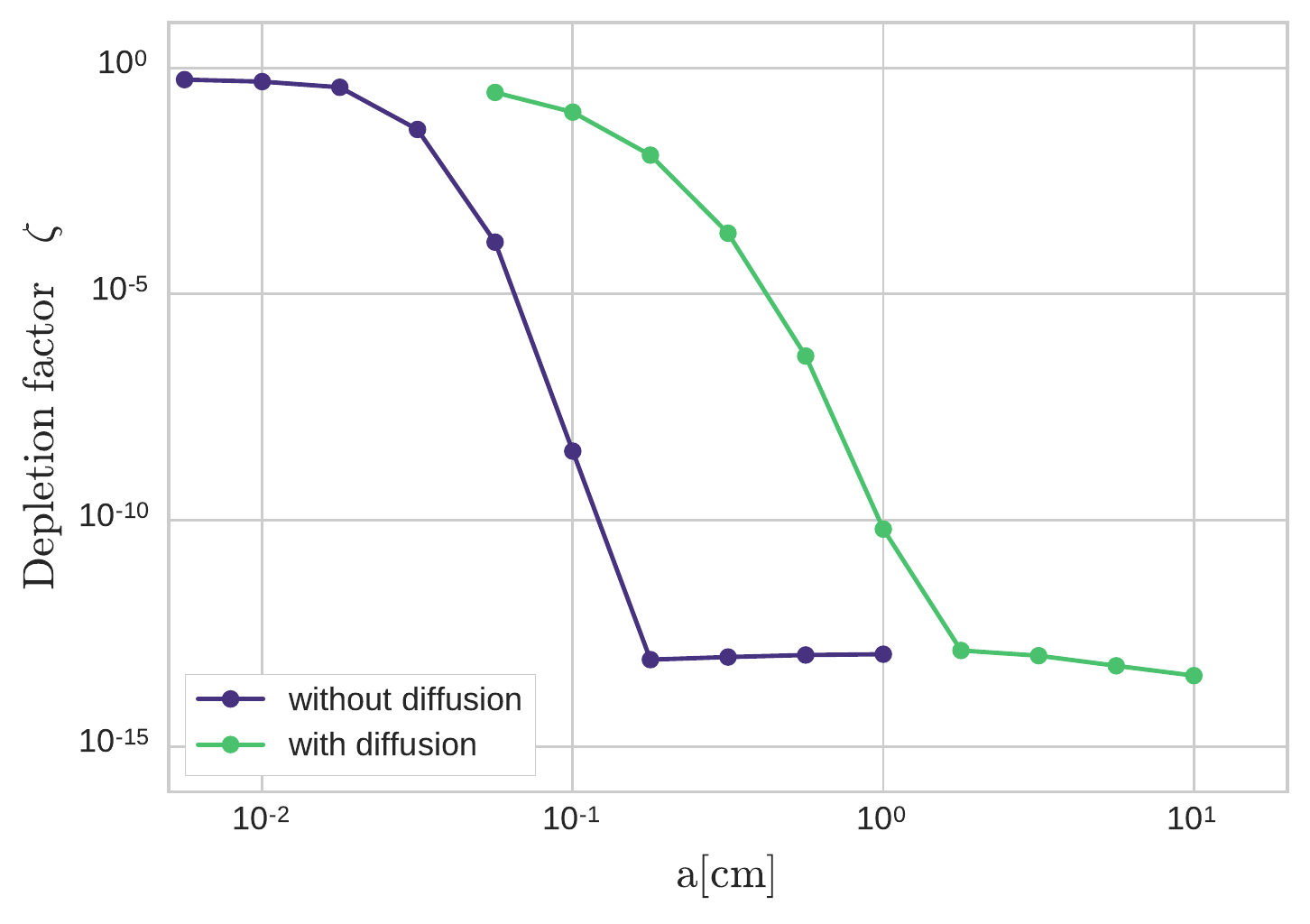}\vspace{-1ex}
\caption{Effective dust permeability of the gap as a function of grain size with and without including the diffusive dust flux.}
\label{fig:difffilter}
\end{figure}

Regarding the depletion of the inner system, in Figure~\ref{fig:difffilter} we find that the effect of the diffusion (in the realm of the fiducial model) is a shift of the filtration cutoff towards larger grain sizes by almost one order of magnitude. This effect can qualitatively be understood by diffusive ``tunneling'' through the dust barrier at the pressure maximum, which however relies on the presence of a non-vanishing gradient in the dust concentration \citep[also cf. the discussion in][section 4.2]{Zhu2012}. Note, however, that the size threshold is not well captured in the simplified one-dimensional model (see left-hand panel of Fig.~\ref{fig:sizediff}). This potentially indicates that \emph{azimuthal} diffusion of dust in the vicinity of the planet may play a significant role, and that a reduction of the problem to one dimension is not feasible in a straightforward manner. When accounting for the effect of dust diffusion, one should, in any case, bear in mind that studies that have tried to quantify the turbulent Schmidt number from direct simulations find a range of values, and also show that diffusion can become anisotropic depending on the circumstances \citep{2015ApJ...801...81Z}.

\subsection{The effect of dust feedback}\label{sec:feedback} 

We have so far ignored the change in gas momentum $-\Sigma_{\rm d}\mathbf{f}_{\rm d}$ in Equation~(\ref{eq:NS-gas}), which is a simple consequence of momentum conservation. We have neglected this effect, arguing that the dust mass is negligible in comparison to the gas mass. This allowed us to consider multiple dust grain sizes simultaneously without defining an initial grain size distribution. However, we have seen that large enough dust grains can become trapped in the outer disk and that this locally enhances the dust-to-gas ratio, $\varepsilon$, possibly by some orders of magnitude. In addition, an ensemble of dust particles can also modify the gas orbital motion and thereby affect the components' mutual dynamics \citep[e.g.][]{Tanaka2005, Okuzumi2012}. In the following, we turn our attention to how feedback might change the behavior of the gap as a particle filter/trap.

As indicated above, when including the feedback, the dust-to-gas ratio is not arbitrary anymore.

We again assume a fiducial value of $\varepsilon=0.01$, as plausibly inherited from the protostellar envelope during PPD formation. Our setup, in some sense, signifies a rather extreme scenario in which all the dust is represented by a single particle size. Alternatively, distributing the combined dust mass into several mass bins would reduce the (initial) dust-to-gas ratio for the the size range in which the Stokes number peaks -- accordingly reducing potential implications caused by the dust feedback onto the gas. In conclusion, the effect of the feedback is exaggerated in our simplified case (using a single dust size), which has the virtue that it provides us with an upper limit of the contribution of the feedback.

\begin{figure*}
\centering
\includegraphics[width=.45\textwidth]{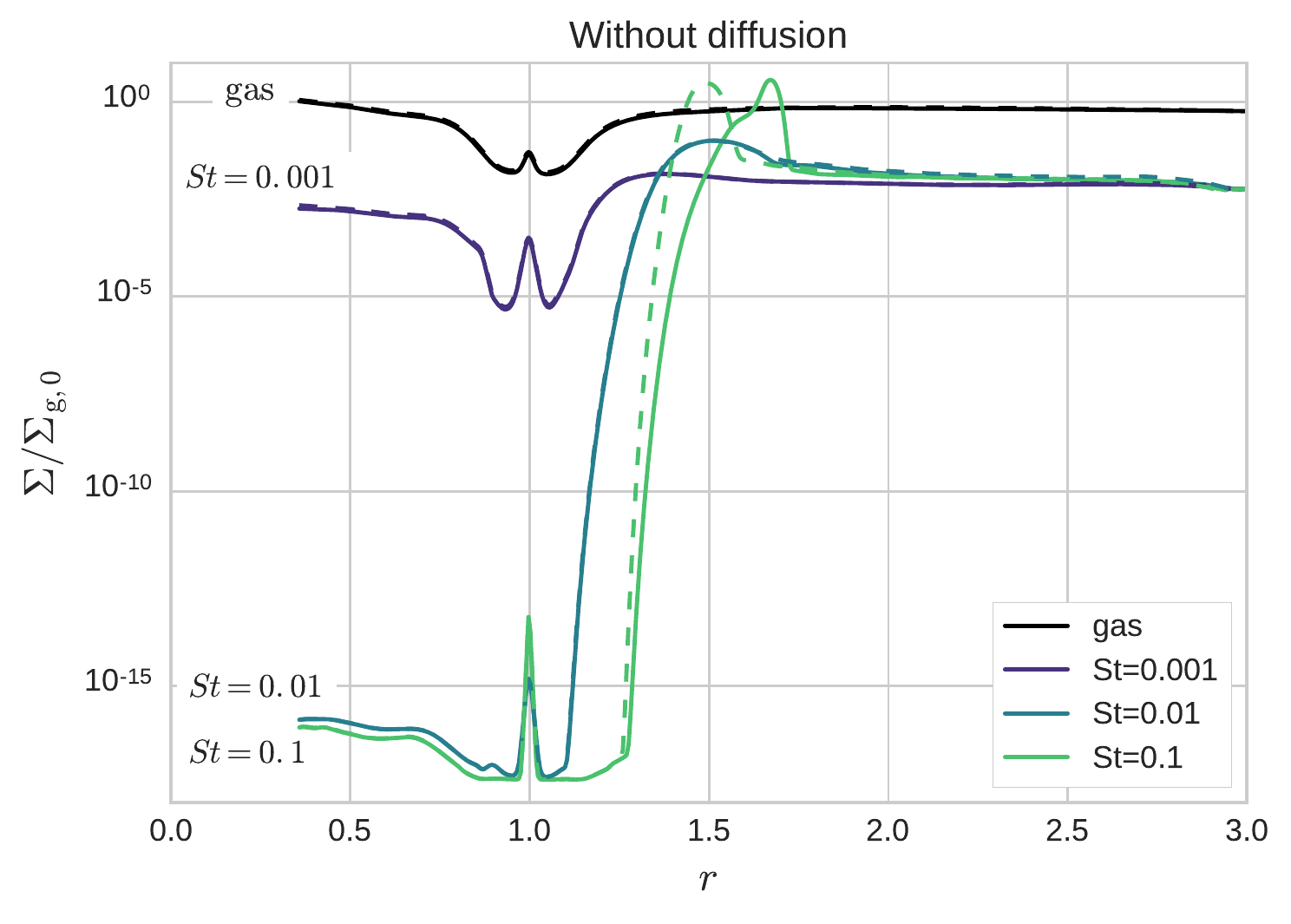}
\includegraphics[width=.45\textwidth]{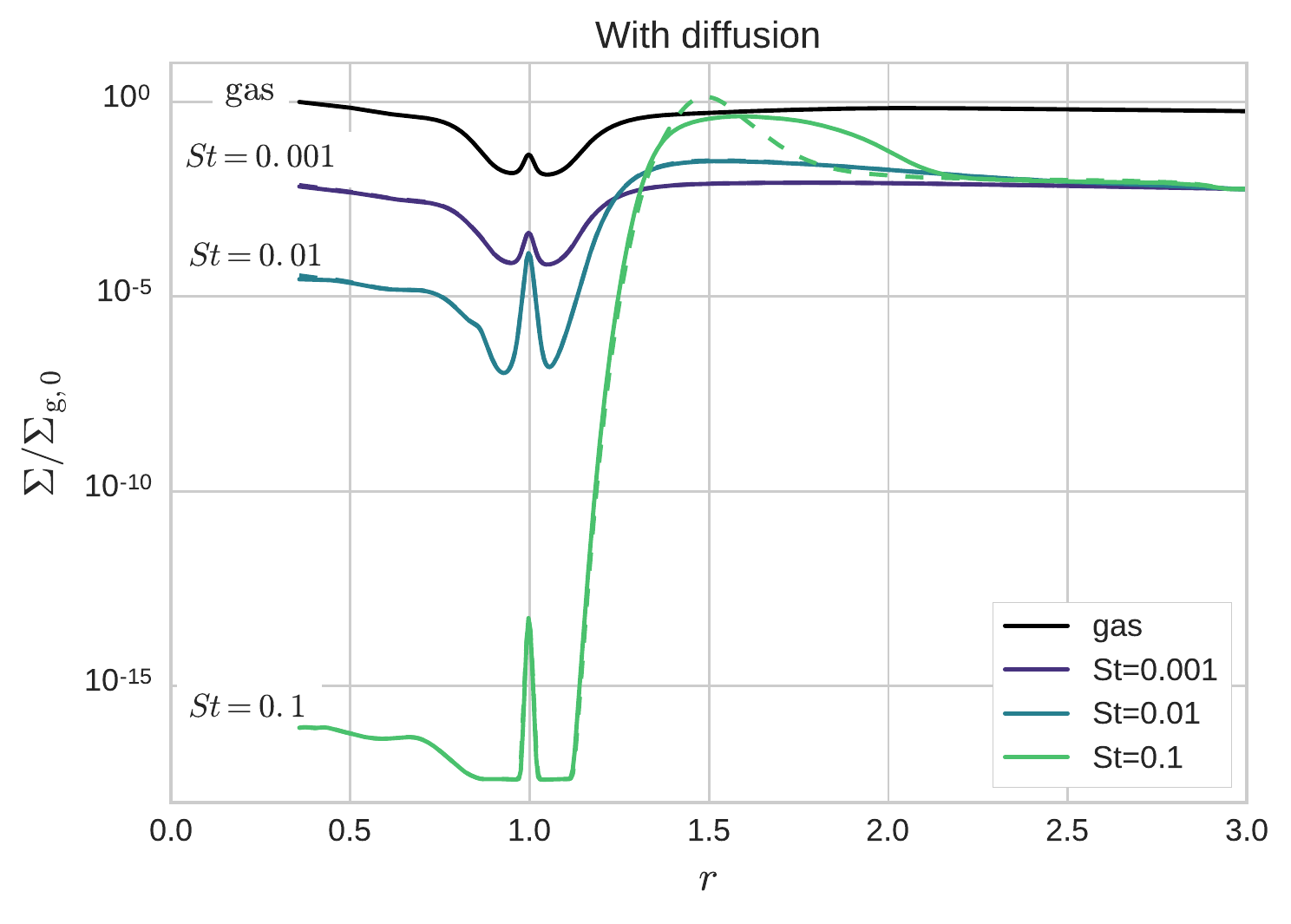}
\caption{Azimuthally averaged surface density after 20,000 planetary orbits, where the dust was introduced to the system after 10,000 planetary orbits. The left and right panels show the case \emph{without} and \emph{with} diffusion. Solid lines show the setup \emph{including} feedback of the dust, the dashed lines without feedback. Every dust species was simulated in a \emph{separate} run.}
\label{fig:feedback}
\end{figure*}

We investigate the effect of the dust feedback for three different (fixed) Stokes numbers. In Figure~\ref{fig:feedback}, we present results for two separate cases, that is, without including the diffusive dust flux (left panel) and when including it (right panel). In both cases, we compare simulations \emph{with} feedback (solid lines) to the outcome in the corresponding simulation \emph{without} feedback (dashed lines). For the two smallest Stokes numbers ($\St=0.001$, $0.01$) we see that including the feedback does not have a significant effect on the final outcome of the dust distribution. This is expected since the dust does not pile up much for such weakly coupled particles. For $\St=0.1$, this is different, since in the region of the pressure maximum the dust density even exceeds the local gas density.

The left panel of Figure~\ref{fig:feedback}, moreover, illustrates that including feedback shifts the position of the maximum of the surface density of the dust towards larger radii. This can be understood by momentum conservation as was for instance shown by \cite{Kanagawa2017}. In regions of a high dust-to-gas ratio, the inwards drifting dust causes the gas to drift outwards as a consequence of momentum transfer. Therefore, the maximum of the gas density is slightly displaced to larger radii, in turn shifting the position of the pressure maximum outwards. With this, the equilibrium point for the dust to collect occurs at a larger radius as well.

In the case of diffusion being included along with feedback, the peak of the distribution is smoothed out even more, so that it becomes more of a plateau than a peak. The reason for this is the same as before. When the dust density equals the density of the gas, it pushes the gas outwards, spreading out the pressure bump and thereby smoothing out the trapping region. We conclude that neglecting the feedback is legitimate when considering the filtration process with the focus of attention on the outcome in the inner region. For the characteristics of the dust trap itself, and for the dust density profile in its vicinity, however, including the feedback can make a non-negligible difference.

\section{Discussion}\label{sec:discussion} 

The numerical study presented in the previous section delineates how the efficiency of size-dependent dust filtration via a planet-induced gap depends on the most central parameters, that is, the planet mass, level of disk turbulence, and inclusion of dust diffusion. Assuming a simplified enhanced viscosity prescription, of course, grossly neglects non-turbulent regions of PPDs as well as potential alternative transport mechanisms \citep[see][for a recent account]{2014prpl.conf..411T}. Also, in the regions where no magnetorotational turbulence ensues, purely hydrodynamic instabilities may arise \citep[e.g.,][]{2013MNRAS.435.2610N,2016A&A...586A..33U}, along with non-trivial consequences for the transport of embedded solids \citep{2016A&A...594A..57S}.
In the following, we discuss further potential limitations of our approach and place the results in context within the current paradigm of planet formation theory.

\paragraph{Validity of the pressureless fluid approach}

As mentioned in section~\ref{sec:dust_disk}, treating the dust as a pressureless fluid is not strictly valid in the limit of large Stokes numbers, and we here briefly assess \emph{a posteriori} whether our models comply with this requirement. In Figure~\ref{fig:2ddens}, it can be seen that for a Jupiter-like planet, the surface density is not reduced by more than a factor $\mathcal{O}(10^{-2})$ anywhere in the gap. This is correctly described by the azimuthal averages presented in, for example, Figure~\ref{fig:gasgap}. Moreover, from looking at Figure~\ref{fig:acc-process}, it becomes clear that the dust tends to follow the gas over-density along the spiral arms -- in this regard, the following estimates can perhaps be considered as safe limits.
More quantitatively, the fiducial model presents an unperturbed surface density of $\sim 900\g\cm^{-2}$, which puts the surface density at the location of the gap at $\sim9\g\cm^{-2}$. Thus, particles with sizes of the order $1\cm$ represent a Stokes number smaller than $\sim 1/2$ everywhere. In this regard, our approach is sensible.
We note, however, that there are two occasions for which the Stokes number does exceed this value inside the gap, namely for centimeter-sized particles in the $q=2\times10^{-3}$ run, and for the $a=10\cm$ particles in the run with diffusion. However, the $q=2\times10^{-3}$ simulation in fact shows that particles with $a=0.1\cm$ are already completely filtered out, and (asserting monotonic behavior) results with larger particles are thus not necessary. Consequently, none of the conclusions in our paper are based on results for Stokes numbers exceeding unity.

\paragraph{Potential effect of coagulation and fragmentation} In our model, we did not include particle--particle interactions between grains. However, depending on the number density of the dust, collisions between grains can be quite important and lead to coagulation and/or fragmentation -- depending on their physical properties and their collisional velocities \citep[e.g.,][]{Birnstiel2010,Drazkowska2014}. If grains that would normally be filtered at the barrier manage to fragment through collisions in the outer parts of the disk, they could eventually be transported through the gap, since their Stokes number is efficiently reduced. We can estimate the grain size of the fragmentation barrier by adopting the formula given in \cite{Birnstiel2009}:
\begin{equation}
a_\mathrm{max} \simeq \frac{ C\,\Sigma_\mathrm{g}}{ \pi \alpha \rho_\mathrm{int}}\frac{v^2_\mathrm{f}}{c^2_\mathrm{s}}\, ,
\end{equation}
where $v_\mathrm{f}$ is the fragmentation velocity, and $C$ is an order unity constant. In our fiducial model a typical value of $a_\mathrm{max}$ inside the pressure bump would be of the order of $10\, \mathrm{cm}$ assuming $v_\mathrm{f} \approx 10\, \mathrm{m}\,\mathrm{s}^{-1}$.

Conversely, small-enough particles that have been transported through the gap, may subsequently grow via dust coagulation. With timescales generally being shorter in the inner system, this effect can potentially replenish the number density of larger-size dust even inside the planet's orbit. As a consequence, adding grain coagulation and fragmentation might substantially alter the size distribution inside the barrier and hence could be an important extension towards a more realistic model.

\paragraph{Potential effect of accretion onto the planet} One process we did not discuss until now is the accretion of dusty gas by the planet itself. As material is being transported through the gap at the azimuthal location of the planet, it will likely encounter the gaseous envelope / circumplanetary disk forming around the planet. \citet{Lissauer2009} find that a giant planet accretes mass even after it has opened a gap which, however, is subject to material being replenished at a sufficient rate \citep{2013ApJ...779...59G,2014Icar..232..266M}. In effect, accretion onto the planet means that the replenishment of the inner system is additionally modulated by a certain factor. \cite{Lubow2006} show by simulating gas accretion onto the planet, that for a planet of Jupiter mass most of the gas (that is, up to 90\%) entering the gap region is accreted by the planet -- potentially reducing the supplied dust mass in the inner system by about an order of magnitude.

\paragraph{Potential effects of planet migration} In all our simulations the planet is not allowed to migrate. Type II migration is the dominant migration mode for giant planets, and the magnitude and sign of this migration can deviate from the typical viscous accretion speed \citep[][]{Duffell2014,Durmann2015}. We can, however, speculate on the possible modifications of our results depending on the sign and magnitude of the migration rate.
If the planet migrates in the \emph{classical} type II regime, where the migration rate is attached to the viscous flow, the relative velocity to very well-coupled dust particles becomes significantly smaller, probably reducing the efficiency of the gap-crossing transport. This, however, does not modify the properties of the pressure trap, allowing larger particles to pile-up since the radial speed is orders of magnitude larger than the viscous speed. The same is valid when the planet migrates inward, faster than the viscous flow but slower than the typically trapped dust particles. If the planet migrates at the type III regime, which is very fast \citep[][]{Masset2003}, even the faster particles could no longer be trapped at the edge of the gap.
Given all the uncertainties on the actual magnitude and sign of migration, it is beyond the scope of this work to study the permeability properties in terms of the planet migration rate.

\paragraph{Consequence for planet formation} The fact that there is no planet bigger than Earth inside Jupiter's orbit in the Solar System is striking in comparison to the known sample of observed extrasolar systems \citep[see][for a comprehensive review]{Winn2015}. From the ever increasing census of exoplanets, it has become clear that so-called super-Earths (sometimes also referred to as mini-Neptunes) are by far the most prevalent type of planet \citep{Batalha2013,Silburt2015}. These planets are not only a fair deal more massive than Earth, but also harbor a significant volatile content in form of icy mantles or gaseous atmospheres. Theoretical considerations indicate that to obtain (and retain during late-stage impacts) such volatile components, the rocky cores of the protoplanets have to grow to several earth masses \citep{2015MNRAS.448.1751I}. There are, however, plausible explanations as to why this type of planets have not formed in the Solar System: \cite{Izidoro2015} propose that the presence of Jupiter would stop the inward migration of any Super-Earth coming from the outer system (where they are more likely to form), which could otherwise potentially end up in the inner system. In an alternative approach, \cite{Morbidelli2015} argue that inside of the snow-line the typical pebble sizes and fluxes from the outer system are simply not favorable to form the massive cores that are capable of forming planets beyond small terrestrial types.

Jupiter acting as a particle barrier could add to this modification of the size distribution of pebbles. First of all, the filtration increases the dust-to-gas ratio just outside of the planet's orbit, which simultaneously reduces the dust-to-gas ratio in the inner system. Furthermore, the grains that pass through unaffected have small Stokes numbers and thus are tightly coupled to the gas, an attribute that prevents them from effectively being subject to the streaming instability \citep{Johansen2009} or pebble accretion \citep{Lambrechts2012,Ormel2017}. Both of these processes, are expected to greatly boost planet formation timescales and hence the efficiency of producing super-Earths. Outside of Jupiter's orbit, the situation is the opposite. \cite{Morbidelli2012} found that for a forming planet exists a so-called \textit{pebble isolation mass}. This mass is reached when the planet becomes heavy enough to shield its orbital region from the inflow of pebbles and by this terminates its phase of pebble accretion, which allows the planet to cool more efficiently. A planet of Jupiter mass is well above this isolation mass and therefore, the pebbles are neither accreted by the planet nor transported to the inner system -- and by this creating an excess in the outer system that is favorable for further planet formation.

\paragraph{Consequence for chondritic measurements} Recent results of isotopic measurements of chondrules \citep[e.g.,][]{Olsen2016,Budde2016,Kruijer2017} suggest the existence of two spatially separated, isotopically distinct reservoirs of material found in the early Solar System. We show in this study that a Jupiter-mass planet comprises a viable barrier to the otherwise rapid inward dust transport effectuated by aerodynamic drag, and that the efficiency of this filtration is intrinsically size-dependent.
\newline

As our final remarks, we want to mention that the discrepancy which was found between the 1D and 2D models arguably motivates a full-blown three-dimensional study of the problem. Including the vertical direction in the simulations will, for instance, allow for height-dependent accretion flows and, moreover, dust settling, which might alter the effective Stokes number (that is, evaluated over a particle's entire passage through the planet/gap system) of a dust population with a given fixed particle radius. As a consequence, the radial dust transport in the presence of the planet may well be affected by complex three-dimensional effects.

\section{Conclusions}\label{sec:conclusion} 

In this paper, we have studied the process of dust transport in viscous gaseous disks in the presence of a gap carved by a giant planet. We have shown that the outer gap-edge acts as a semipermeable filter, retaining sufficiently aerodynamically decoupled particles in the outer system, and modulating the size distribution of dust grains prevalent in the inner system.

The transport of dust through a gap produced by a giant planet is highly dependent on dust size. The exact grain size for the transition depends on such parameters as the turbulent viscosity, the planet mass and the stellar accretion rate. The filtration becomes gradually more efficient for larger particles and accordingly the transition from low to high permeability is not an abrupt function but gradually changes over about one order of magnitude in particle radius.

We remark that there is a discrepancy between results obtained from a 1D and a 2D treatment, which is due to the non-axisymmetric nature of the planet's gravitational potential and the subtle effects it has on the coupled dynamics of the gas and dust. With 2D simulations being readily available on accelerated architectures, we caution against the use of a simplified 1D description. Moreover, we highlight the difference of studying dust species with a fixed (or for a more realistic scenario, evolving) particle radius, compared to a fixed Stokes number. We point out that the fundamental change of character of the aerodynamic coupling strength when entering the low-density gap region is not captured by simply assuming a fixed Stokes number. In addition, we found that using fixed particle sizes significantly reduces the peak dust-to-gas ratio of particles piling up at the outer gap edge.

\acknowledgments
We thank Troels Haugb{\o}lle, Daniel Wielandt, and Martin Bizzarro for many useful discussions, and Cornelis Dullemond for providing comments on an earlier draft of this manuscript, as well as the anonymous referee for a timely and useful report.
We are grateful for the opportunity to participate in the annual retreat of the Center for Star and Planet Formation, at the Natural History Museum of Denamrk, where the ideas leading to this work were shaped.
This project has received funding from the European Union's Horizon 2020 research and innovation programme under grant agreement No 748544 (PBLL).
The research leading to these results has received funding from the European Research Council (ERC) under the European Union's Horizon 2020 research and innovation programme (grant agreement No 638596) (OG).
The research leading to these results has received funding from the European Research Council (ERC) under the European Union's Seventh Framework programme (FP/2007-2013) under ERC grant agreement No 306614 (MEP).
This research was supported by the Munich Institute for Astro- and Particle Physics (MIAPP) of the DFG cluster of excellence ``Origin and Structure of the Universe''.
Computations were performed on the \texttt{astro\_gpu} partition of the Steno cluster at the HPC center of the University of Copenhagen.
\software{This work has made use of IPython \citep{Ipython}, NumPy \citep{Numpy} and Matplotlib \citep{Matplotlib} for creating figures.}

        \appendix
    \section{Dimensionless formulation of the equations}\label{ap:Appendix}

We briefly outline here how to consider this problem in terms of dimensionless variables. This enables us to find the corresponding scaling (for instance, for the physical grain size at which the efficient filtering sets in), when considering a protoplanetary system that is different in terms of stellar mass, $M_\ast$, or planet location, $r_\mathrm{P}$. The new results can be obtained by merely rescaling the variables accordingly without the need to perform new simulations. In order to define dimensionless variables we consider the length-scale provided by $r_0$ and
the time-scale $t_0$ provided by the inverse of the angular frequency $\Omega_0 = \sqrt{GM_\ast/r_0^3}$. These induce natural scales for the velocity $v_0 = \sqrt{GM_\ast/r_0}$ and the surface density $\Sigma_0 = \dot{M}/\left(3\pi \alpha h_0^2 \Omega_0 r_0^2 \right)$.

The equations of interest, viz. the continuity equations and momentum equations for gas and dust, may be then written in full dimensionless form in terms of the dimensionless differential operators $\nabla' = r_0\, \nabla$ and $\partial t' = \Omega_0\, \partial t$ as
\begin{eqnarray}
\frac{\partial \Sigma_{\rm g}'}{\partial t'} + \nabla'\cdot \left(\Sigma_{\rm g}' \mathbf{v}' \right) &=& 0, \label{eq:A1}\\
\frac{\partial \Sigma_{\rm d}'}{\partial t'} + \nabla'\cdot \left(\Sigma_{\rm d}' \mathbf{v}' + \mathbf{j}'\right) &=& 0, \\
\displaystyle{\frac{\partial \mathbf{u}'}{\partial t'} + \left(\mathbf{u}'\cdot \nabla'\right)\mathbf{u}'} &=& \displaystyle{ - \frac{\nabla' P'}{\Sigma_{\rm g}'} - \nabla'\phi' -  \frac{\nabla' \cdot \tau'}{\Sigma_{\rm g}'} - \frac{\Omega_{\rm K}'\Sigma_{\rm g}'}{{\rm St}_0}\left( \mathbf{u}' - \mathbf{v}'\right)}, \\
\displaystyle{\frac{\partial \mathbf{v}'}{\partial t'} + \left(\mathbf{v}'\cdot \nabla'\right)\mathbf{v}'} &=& \displaystyle{- \nabla'\phi' + \varepsilon \frac{\Omega_{\rm K}'\Sigma_{\rm g}'}{{\rm St}_0}\left( \mathbf{u}' - \mathbf{v}'\right)} \label{eq:A4}\,,
\end{eqnarray}
where we have to substitute the abbreviation
\begin{equation}
\tau' \equiv \Sigma'_{\rm g} \nu' \left[ \mathbf{\nabla}' \mathbf{u}' + (\mathbf{\nabla}'\mathbf{u}')^T - \frac{2}{3}(\mathbf{\nabla}'\cdot \mathbf{u}')\mathit{\mathbf{I}}\right]\,.
\end{equation}
Here, the complete set of dimensionless variables is given by
\begin{align}
t' &= t\,/t_0 \,, &
r' &= r\,/r_0 \,, &
\Omega_{\rm K}' &= \Omega_{\rm K}/\Omega_0 \,, \nonumber\\
\mathbf{u}' &= \mathbf{u}/v_0\,,  &
\mathbf{v}' &= \mathbf{v}/v_0\,,  &
P' &= P/(v_0^2\Sigma_{0})\,, \nonumber\\
\Sigma'_{\rm g,d} &= \Sigma_{\rm g,d}/\Sigma_0\,, &
{\rm St}_0 &= a_0 \rho_{\rm int} \pi/\left(2 \Sigma_0\right) \,, &
\nu' &= \nu/(v_0r_0) \,, \nonumber\\
q &= m_{\rm P}/M_\ast\,,  &
\phi' &= -\frac{1}{r'} - \frac{q}{\sqrt{r'^2 \!+\! 1 \!-\! 2r'\cos{\varphi}}} + qr'\cos{\varphi} \,.
\end{align}
In the equations of motion written in their dimensionless form, that is, Equations~(\ref{eq:A1})--(\ref{eq:A4}), one can identify five dimensionless parameters that define the problem in an unambiguous fashion, namely $\alpha$, $h_0$, $q$, ${\rm St}_0$, $\varepsilon$. All our results can then be rescaled simply by demanding that the five parameters above remain unchanged.

For instance, keeping $\mathrm{St}_0$ fixed implies, that if one was to change the background density, $\Sigma_0$, the particle size, $a$, would have to change in a proportional manner. From this we derive a scaling rule for the particle size when changing model parameters $M_\ast \rightarrow \widehat{M}_\ast$, $r_0 \rightarrow \widehat{r}_0$ and $\dot{M}\rightarrow\widehat{\dot{M}}$:
\begin{equation}
\frac{\widehat{a}_0}{a_0} = \frac{\widehat{\Sigma}_0}{\Sigma_0} = \frac{\widehat{\dot{M}}}{\dot{M}} \sqrt{ \frac{M_\ast}{\widehat{M}_\ast} \frac{r_0}{\widehat{r}_0}}
\label{eq:shift}
\end{equation}
We want to illustrate this with an explicit example. All the results given in this paper were for a Solar System model with $M_\ast = 1\,\mathrm{M}_\odot$, $r_0 = 5.2\,\mathrm{AU}$ and $\dot{M}=10^{-7}\,\mathrm{M}_\odot\,\mathrm{yr}^{-1}$. Let us assume we are interested in a smaller and older system with a more massive star ($\widehat{M}_\ast=2\,\mathrm{M}_\odot$, $\widehat{r}_0= 1\,\mathrm{AU}$, $\widehat{\dot{M}}  = 10^{-8}\,\mathrm{M}_\odot\,\mathrm{yr}^{-1}$). Then, inserting these values into Equation~(\ref{eq:shift}) tells us, that a dust grain of size $a=1\,\mathrm{cm}$ in our results would correspond to a size of $\hat{a}\approx 0.16\,\mathrm{cm}$ in the modified system.

\end{document}